\documentclass[acmsmall,screen]{acmart}
\usepackage{amsmath} 
\usepackage{graphicx}
\usepackage{booktabs,tabularx}
\usepackage{xcolor}
\usepackage{soul}
\usepackage{colortbl}
\graphicspath{{}}

\definecolor{lightgray}{gray}{0.95}

\setcopyright{acmlicensed}
\acmDOI{}

\usepackage{xspace}

\makeatletter
\DeclareRobustCommand\onedot{\futurelet\@let@token\@onedot}
\def\@onedot{\ifx\@let@token.\else.\null\fi\xspace}

\newcommand{\rqone}{Can probability-based UQ serve as a general performance indicator in within-project defect prediction?\xspace}
\newcommand{\rqtwo}{Does high predictive performance imply well-calibrated probabilities in within-project defect prediction?\xspace}
\newcommand{\rqthree}{Do relationships among UQ, performance, and calibration persist under cross-project defect prediction?\xspace}

\title{Understanding Software Defect Prediction: A Large-scale Empirical Study Across Uncertainty Quantification and Performance Evaluation
}

\author{Ranjun Peng}
\email{1230024861@student.must.edu.mo}
\orcid{0009-0004-8469-0228}
\affiliation{
  \institution{School of Computer Science and Engineering, Macau University of Science and Technology}
  \city{Macao SAR}
  \postcode{999078}
  \country{China}
}

\author{Xuan Xie}
\authornote{Corresponding author.}
\email{xiexuan@must.edu.mo}
\orcid{0000-0003-3981-8515}
\affiliation{
  \institution{School of Computer Science and Engineering, Macau University of Science and Technology}
  \city{Macao SAR}
  \country{China}
  \postcode{999078}
}

\author{Rubing Huang}
\email{rbhuang@must.edu.mo}
\orcid{0000-0002-1769-6126}
\affiliation{
  \institution{School of Computer Science and Engineering, Macau University of Science and Technology}
  \city{Macao SAR}
  \postcode{999078}
  \country{China}  
}

\author{Wenbin He}
\email{bobo_hewenbin@163.com}
\orcid{0000-0001-5253-4651}
\affiliation{
  \institution{State Key Laboratory of Intelligent Manufacturing Equipment and Technology, School of Mechanical Science and Engineering, Huazhong University of Science and Technology}
  \city{Wuhan}
  \country{China}
  \postcode{430074}
}

\author{Zhijie Wang}
\email{zhijie.wang@concordia.ca}
\orcid{0000-0003-4559-5426}
\affiliation{
  \institution{Department of Computer Science and Software Engineering, Concordia University}
  \city{Montreal}
  \state{Quebec}
  \country{Canada}
  \postcode{H3G 1M8}
}

\newenvironment{compactitem}
  {\begin{itemize}
   \setlength{\itemsep}{0pt}
   \setlength{\parsep}{0pt}
   \setlength{\topsep}{2pt}
   \setlength{\partopsep}{0pt}}
  {\end{itemize}}

\definecolor{boxgray}{gray}{0.92}
\newsavebox{\findingboxcontent}
\newenvironment{findingbox}
  {\par\addvspace{0.5\baselineskip}\noindent\begingroup
   \setlength{\fboxrule}{0.8pt}
   \setlength{\fboxsep}{4pt}
   \begin{lrbox}{\findingboxcontent}
   \begin{minipage}{\dimexpr\linewidth-2\fboxsep-2\fboxrule-6pt\relax}}
  {\end{minipage}
   \end{lrbox}
   \fcolorbox{gray!60}{boxgray}{\usebox{\findingboxcontent}}
   \par\endgroup\addvspace{0.5\baselineskip}}

\usepackage{pdflscape}
\begin{document}

\begin{abstract}
Software defect prediction (SDP) classifiers produce probabilities used for inspection prioritization, threshold tuning, and risk communication.
Probability-based uncertainty quantification (UQ) characterizes prediction confidence, but whether common UQ metrics reliably indicate performance and calibration remains unclear.
We conducted a large-scale empirical study of probability-based UQ for SDP.
We evaluated five UQ metrics, six performance metrics, and three calibration metrics for 16 representative classifiers.
We analyzed these relationships under two prediction settings: within-project defect prediction (WPDP), using 36 benchmark datasets, and cross-project defect prediction (CPDP), using 32 feature-compatible datasets.
Results showed that UQ was highly context-dependent.
Under WPDP, UQ correlated more consistently with false positive rate and AUC than with MCC, F1 score, and other metrics; these correlations also varied across classifier categories and dataset collections.
Performance and calibration were related but not interchangeable; classifiers with strong discrimination could still exhibit large calibration error.
Under CPDP, several UQ-performance and UQ-calibration correlations weakened or reversed, indicating that uncertainty signals do not reliably transfer across projects.
Thus, UQ should be evaluated against specific performance objectives.
Calibration should be assessed independently using multiple metrics.
Transferred probabilities should be revalidated before guiding quality-assurance decisions.
\end{abstract}



\maketitle

\section{Introduction}
\label{sec:introduction}
Software defect prediction (SDP) aims to identify defect-prone modules before release, so that quality-assurance effort can be focused where it is most likely to improve software quality.
This task has been studied for more than two decades, from early work on static code attributes to large classifier benchmarks and systematic reviews~\cite{menzies2007data,menzies2010defect,lessmann2008benchmarking,li2018progress,matloob2021software,catal2009systematic,hall2011systematic}.
Recent work has expanded SDP through deep representations, attention-based feature fusion, and cross-project learning~\cite{omri2020deep,nam2015heterogeneous,ye2025software,qiu2025features,liu2024sedpgk}.
Nevertheless, metric-based tabular SDP remains the most popular, as a significant number of public benchmarks and industrial pipelines still describe modules using product, process, and change metrics~\cite{nagappan2005use,moser2008comparative,zhao2023systematic,hassan2009predicting}.

SDP is commonly formulated as a binary classification task, where a model predicts whether a software module is defective or clean. 
To make this prediction, learning-based SDP models typically estimate the probability that each module contains defects and then classify modules based on a decision threshold. 
A higher predicted probability indicates a greater likelihood that the module is defective. 
Beyond binary classification, these probabilities are often used for other practical purposes, such as prioritizing modules for inspection, adjusting inspection thresholds to match available resources~\cite{canfora2015defect}, combining predictions from multiple models, and communicating defect risk to developers~\cite{yu2024improving}. 
However, predicted probabilities alone may not fully capture the reliability of a prediction. 
Two modules may receive similar defect probabilities while differing substantially in the confidence that the model has in those estimates. 
To address this limitation, uncertainty quantification (UQ) has been introduced to characterize the ambiguity associated with model predictions~\cite{el2010foundations}. 
By providing an additional measure of confidence, UQ has the potential to support more informed decision-making in SDP. 

Over the last decade, the machine learning community has developed a series of UQ techniques, including probability-based measures, Bayesian approximations, and ensemble-based approaches~\cite{abdar2021review,gal2016dropout,lakshminarayanan2017simple,he2025survey,hullermeier2021aleatoric,hendrycks2016baseline,ovadia2019can}.
Among the various UQ approaches, probability-based uncertainty measures are particularly attractive because they can be derived directly from the predicted class probabilities produced by most SDP models, requiring no changes to the underlying learning algorithm. 
Common examples include confidence, margin, and entropy-based measures, all of which aim to quantify how certain a model is about its prediction.
These measures have also recently been adopted in software engineering tasks such as test selection and automated program repair~\cite{ma2021test,yang2025revisiting,feng2020deepgini,weiss2022simple,xia2023automated}. 
More recently, defect prediction researchers have investigated techniques such as conformal prediction and probability calibration in just-in-time defect prediction settings~\cite{shahini2024empirical,ju2025jit,shahini2025calibration}. 
However, despite the growing interest in uncertainty-aware defect prediction, there remains limited empirical evidence regarding \textit{which probability-based uncertainty measures are most effective for SDP}. 
Furthermore, practitioners lack clear guidance on \textit{how uncertainty scores should be interpreted and incorporated into defect prediction workflows}.

Answering these questions is challenging because SDP classifiers generate probability estimates through fundamentally different mechanisms; for example, probabilities may be derived from likelihood assumptions, sigmoid-transformed margins, neighborhood counts, tree leaf frequencies, softmax activations, or ensemble aggregation.
As a result, the same uncertainty measure may behave differently across classifiers and data distributions. 
This challenge is particularly pronounced in SDP, where datasets often exhibit substantial class imbalance, class overlap, label noise, and project-specific feature characteristics~\cite{agrawal2018better,gong2022comprehensive,wan2024data}.
The challenge is further compounded by differences in feature distributions, defect rates, and labeling practices in cross-project defect prediction (CPDP)~\cite{zimmermann2009cross,nam2013transfer,ovadia2019can}.
It remains unclear whether probability-based uncertainty measures provide consistent and reliable signals across different defect prediction settings.

To bridge the gap, we conducted a large-scale empirical study to examine whether probability-based UQ can be reliably used as empirical evidence for SDP. Specifically, we investigated the following research questions.

\begin{compactitem}
    \item[RQ1] \rqone
    \item[RQ2] \rqtwo
    \item[RQ3] \rqthree
\end{compactitem}

To answer these questions, we collected 40 candidate SDP classifiers and selected 16 behaviorally representative classifiers using classifier output difference clustering~\cite{kuncheva2003measures,bowes2018software}.
Our evaluation spanned 36 datasets from AEEEM, NASA, PROMISE, and ReLink~\cite{d2010extensive,shepperd2013data,jureczko2010towards,wu2011relink}.
For each classifier–dataset pair, we computed five probability-based UQ metrics, six predictive performance metrics, and three calibration metrics.
We then investigated the relationships among uncertainty, predictive performance, and calibration under two prediction settings: within-project defect prediction (WPDP), covering all 36 datasets, and cross-project defect prediction (CPDP), covering 32 feature-compatible datasets.

The study yielded three key observations that challenge common assumptions in uncertainty-aware SDP evaluation. The first two concern WPDP. First, probability-based uncertainty metrics were informative for only a narrow subset of performance measures, most notably false positive rate (FPR) and AUC; for commonly reported metrics such as F1 score, the observed relationships were weak or negligible. Second, predictive performance and probability calibration were related but not interchangeable, as classifiers with strong predictive performance could still produce poorly calibrated probability estimates, limiting the reliability of threshold-based decision making. The third observation concerns the transfer from WPDP to CPDP: relationships between uncertainty and predictive quality were not stable across the two settings, and several relationships observed under WPDP weakened or even reversed under CPDP, suggesting that uncertainty evidence obtained within a project cannot be assumed to generalize to other projects.

In summary, this paper makes the following contributions:
\begin{compactitem}

    \item We conducted a large-scale empirical study of probability-based uncertainty quantification (UQ) for software defect prediction (SDP), covering 16 representative classifiers and 36 benchmark datasets under both WPDP and CPDP settings.
    
    \item We systematically characterized the relationships between probability-based uncertainty metrics, predictive performance, and probability calibration across classifier categories, dataset collections, and prediction granularities.

    \item We identified the strengths, limitations, and practical applicability of probability-based UQ methods in SDP, providing evidence-based guidance for researchers and practitioners on selecting and interpreting uncertainty metrics.

\end{compactitem}

The remainder of the paper is organized as follows.
Section~\ref{sec:related_work} reviews defect prediction, dataset and transfer challenges, and probabilistic predictive evaluation.
Section~\ref{sec:study_design} describes the study design and protocol.
Sections~\ref{sec:rq1}--\ref{sec:rq3_cpdp} report the results for the three research questions.
Section~\ref{sec:discussion} discusses the implications of our findings for uncertainty-aware SDP evaluation and reporting.

\section{Related Work}
\label{sec:related_work}
This study drew on three lines of work: defect prediction classifiers, dataset and transfer challenges, and probabilistic predictive evaluation.
Prior SDP research has examined benchmark classifiers, transfer learning, class imbalance, and project-specific data distributions.
UQ research has proposed measures of predictive uncertainty, but most evidence comes from general machine learning and deep learning.
Prior work has shown that defect prediction performance is sensitive to classifier choice and dataset differences, and that probability estimates require evaluation separately from classification performance.
What remains unresolved is whether probability-based UQ metrics from heterogeneous classical classifiers provide consistent performance indicators in metric-based module-level SDP.
This section reviews the evidence behind that gap.

\subsection{Software Defect Prediction}
\label{subsec:sdp}

Software defect prediction aims to identify defect-prone modules before release so that inspection and testing effort can be allocated to modules with higher defect risk~\cite{zhao2023systematic}.
Many SDP studies formulate the task as binary classification using product metrics, process metrics, or both.
Common classifiers in these studies include Logistic Regression, Naive Bayes, Support Vector Machines, Decision Trees, and ensemble learners~\cite{lessmann2008benchmarking,tantithamthavorn2016empirical,matloob2021software,laradji2015software}.
Process metrics such as code churn and developer activity also improve defect prediction~\cite{nagappan2005use,moser2008comparative}.
When a project has little labeled history, cross-project defect prediction transfers training data from other projects~\cite{zimmermann2009cross,nam2013transfer,nam2015heterogeneous}.
Work on semantic code representations has expanded, but metric-based tabular SDP remains relevant because many public and industrial datasets use software metrics~\cite{omri2020deep,zhao2023systematic,ye2025software,qiu2025features}.

Benchmark studies have shown that classifier evaluation in SDP is sensitive to validation protocol, parameter tuning, and performance metric choice~\cite{ghotra2015revisiting,lessmann2008benchmarking,tantithamthavorn2016automated,tantithamthavorn2018impact}.
This matters for UQ because uncertainty scores are derived from the same probability outputs produced by these classifiers.
These studies have shown that classifier choice and validation design shape reported performance.
They have left open whether probability outputs from these classifiers yield uncertainty scores that track performance or calibration.

\subsection{Dataset and Transfer Challenges in Software Defect Prediction}
\label{sec:rw_dc}

Defect data contain class imbalance, class overlap, label noise~\cite{kim2011dealing}, and project-specific feature distributions~\cite{shepperd2013data,gong2022comprehensive,wan2024data,fan2019impact}.
These properties affect both classification performance and probability estimates.
Class imbalance can bias the learned decision boundary toward the clean class, while overlap and noise make some modules difficult to classify~\cite{smith2014instance,lorena2019complex}.

SDP studies have used class imbalance treatments~\cite{wang2013using,tantithamthavorn2018impact,goyal2025current}.
Random undersampling, SMOTE, Borderline SMOTE, and ADASYN change the training distribution to improve minority class detection~\cite{elhassan2016classification,han2005borderline,he2008adasyn}.
However, imbalance treatment does not remove all sources of dataset heterogeneity.
Prior work has shown that overlap and noisy labels can remain relevant after imbalance correction~\cite{agrawal2018better,gong2022comprehensive,wan2024data}.
Oversampling can introduce synthetic points in ambiguous regions, which may affect probability estimates~\cite{blagus2013smote,agrawal2018better}.
These data challenges motivate an evaluation of whether probability-based uncertainty behaves consistently across dataset collections, prediction granularities, and cross-project transfer settings.

\subsection{Uncertainty Quantification}
\label{sec:rw_uq}

Uncertainty quantification characterizes ambiguity in classifier outputs~\cite{abdar2021review,he2025survey,hullermeier2021aleatoric}.
In this paper, UQ refers to uncertainty metrics derived from predictive probabilities.
In software engineering, uncertainty-aware methods support test selection and automated program repair, where entropy or confidence scores can help prioritize inputs or candidate patches~\cite{ma2021test,yang2025revisiting}.

Calibration is separate from predictive uncertainty.
A classifier is calibrated when predicted probabilities match empirical outcome frequencies~\cite{niculescu2005predicting,guo2017calibration,nixon2019measuring}.
Calibration has been studied for defect prediction and imbalanced classification settings~\cite{dal2015calibrating,van2022harm}.
However, high classification performance does not ensure calibrated probabilities.
This is relevant for SDP because predicted probabilities may support threshold setting, inspection prioritization, and cost-sensitive decisions.

Recent work has examined uncertainty-aware prediction in just-in-time defect prediction.
Shahini et al.~studied conformal prediction for just-in-time defect prediction and evaluated whether conformal prediction can identify predictions that are likely to be incorrect~\cite{shahini2024empirical}.
Related work has also examined calibration in just-in-time defect prediction and shown that probability-based decision support requires explicit calibration assessment~\cite{shahini2025calibration}.
These studies have addressed predictive confidence in change-level just-in-time settings.
They have not answered whether probability-based uncertainty from heterogeneous classical classifiers serves as a general signal for performance, calibration, or cross-project transfer in module-level SDP.

Three empirical gaps remain.
First, prior work has not established whether probability-based UQ metrics correlate with predictive performance across the heterogeneous classifiers and data conditions in SDP.
Second, it has not been shown whether high predictive performance implies well-calibrated probabilities in SDP.
Third, it is unknown whether UQ--performance and UQ--calibration correlations persist under cross-project transfer.
We addressed these gaps by evaluating correlations between UQ and performance and between UQ and calibration across classifiers, dataset collections, and prediction settings.

\section{Empirical Study of Uncertainty, Predictive Performance, and Calibration}
\label{sec:empirical_study}

Building on the three knowledge gaps identified in Section~\ref{sec:rw_uq}, this empirical study treated probability-based UQ as a measurable signal rather than as an assumed indicator of model quality.
The study examined whether uncertainty scores derived from classifier probability outputs aligned with predictive performance, calibration, and transfer behavior in SDP.
We used a common evaluation protocol across classifiers and datasets, which allowed UQ behavior to be compared across classifier categories, dataset collections, prediction granularities, and prediction settings.
In this study, uncertainty denotes ambiguity in the classifier probability distribution for an individual software module.
A prediction close to an even split, such as 0.51 versus 0.49, is more uncertain than a concentrated prediction.

\subsection{Study Motivation and Research Questions}
\label{sec:study_motivation}

Three knowledge gaps motivate the research questions.
Each gap identifies a specific aspect of probability-based UQ in SDP that lacks systematic empirical evidence.

\noindent\textbf{Gap~1: It remains unclear whether probability-based UQ correlates with predictive performance across heterogeneous classifiers and datasets.}
Classical SDP benchmarks employ a wider set of classifiers than typical UQ studies, including Logistic Regression, Decision Tree, Naive Bayes, Support Vector Machine, Random Forest, and KNN~\cite{lessmann2008benchmarking,tantithamthavorn2016empirical}.
These classifiers estimate probabilities through fundamentally different mechanisms, such as likelihood assumptions, link functions, leaf frequencies, neighborhood counts, margin transformations, or ensemble aggregation~\cite{niculescu2005predicting}.
SDP datasets further vary in class imbalance~\cite{moser2008comparative,wang2013using} and prediction granularity (file-, class-, and method-level).
Prior UQ research has focused predominantly on deep learning~\cite{gal2016dropout,lakshminarayanan2017simple,abdar2021review}, and it has not established whether probability-based UQ provides a consistent performance signal across the diverse classifiers and data conditions found in SDP.
This gap motivates \textbf{RQ1}.

\noindent\textbf{Gap~2: High predictive performance has not been shown to imply well-calibrated probabilities in SDP.}
Predictive performance and probability calibration measure different properties of classifier outputs.
A classifier can separate defective modules effectively while assigning probability values that do not match empirical defect frequencies~\cite{niculescu2005predicting,guo2017calibration}.
Conversely, a classifier may exhibit low aggregate calibration error while providing weak discrimination for a target metric such as MCC or AUC.
These properties support different decisions: discrimination guides inspection ranking, whereas calibration matters when probability values are interpreted as defect likelihoods.
Although prior work has studied calibration in general machine learning~\cite{guo2017calibration,nixon2019measuring} and just-in-time defect prediction~\cite{shahini2025calibration}, systematic evidence on whether high-performing SDP classifiers also produce well-calibrated probabilities remains lacking.
This gap motivates \textbf{RQ2}.

\noindent\textbf{Gap~3: It is unknown whether UQ--performance and UQ--calibration correlations observed within a project persist under cross-project transfer.}
CPDP introduces distribution shifts in feature distributions, defect rates, and labeling practices between source and target projects~\cite{zimmermann2009cross,nam2013transfer,ovadia2019can}.
Class imbalance treatments such as SMOTE further alter the training distribution and may affect probability estimates~\cite{dal2015calibrating,van2022harm}.
These shifts can change both the classifier probability outputs and the data conditions under which UQ metrics are computed.
However, no prior study has examined whether the UQ correlations established in within-project settings remain stable after cross-project transfer.
This gap motivates \textbf{RQ3}.

These three gaps together call for a systematic evaluation of probability-based UQ in SDP.
Accordingly, the study is organized around three research questions:
\begin{compactitem}
    \item \textbf{RQ1:} \rqone
    \item \textbf{RQ2:} \rqtwo
    \item \textbf{RQ3:} \rqthree
\end{compactitem}

\subsection{Study Design}
\label{sec:study_design}

\subsubsection{Benchmark Datasets}

The benchmark contains 36 defect prediction datasets from four established repositories: AEEEM~\cite{d2010extensive}, NASA~\cite{shepperd2013data}, PROMISE~\cite{jureczko2010towards}, and ReLink~\cite{wu2011relink}.
We adopted the same curated dataset suite as Wan et al.~\cite{wan2024data}.
The benchmark comprises 5 AEEEM datasets, 12 NASA datasets, 15 PROMISE datasets, and 4 ReLink datasets.
These corpora are common in defect prediction research~\cite{li2020understanding,tantithamthavorn2016automated,tantithamthavorn2018impact}.
They span different prediction granularities: AEEEM and PROMISE are treated as class-level datasets, NASA as function/method-level datasets, and ReLink as file-level datasets.
Prediction granularity refers to the structural level of the software artifacts being analyzed, such as files, classes, or individual functions.
Because granularity is coupled with dataset collection in this benchmark, granularity-level analyses are interpreted as granularity-aware collection comparisons rather than as evidence of a pure granularity effect.

\subsubsection{Candidate Classifiers}

We constructed the candidate classifier pool from 40 classifiers.
Following the taxonomy in prior studies~\cite{wan2024data}, we organized these candidates into the following groups:

\begin{compactitem}
    \item \textbf{Probabilistic Model:} This group includes \textit{Naive Bayes}, \textit{Complement Naive Bayes}, \textit{Gaussian Process Classifier}, and \textit{Quadratic Discriminant Analysis}.

    \item \textbf{Linear Model:} This group includes \textit{Ridge Classifier}, \textit{Logistic Regression}, \textit{Linear Support Vector Classifier}, \textit{Stochastic Gradient Descent Classifier}, and \textit{Passive Aggressive Classifier}.

    \item \textbf{Instance-Based and Kernel Method:} This group includes \textit{KNN}, \textit{Linear Support Vector Machine}, and \textit{Radial Basis Function Kernel Support Vector Machine}.

    \item \textbf{Tree- and Rule-Based Method:} This group includes \textit{Classification and Regression Tree}, \textit{Boosted Rule Set}, and \textit{Greedy Rule List}.

    \item \textbf{Neural Network:} This group includes two multilayer perceptrons: \textit{One Hidden Layer Multilayer Perceptron} and \textit{Two Hidden Layer Multilayer Perceptron}.

    \item \textbf{General Ensemble Method:} This group includes \textit{Random Forest}, \textit{Extremely Randomized Trees}, \textit{Gradient Boosted Decision Trees}, \textit{Voting Classifier}, \textit{Extreme Gradient Boosting}, \textit{Light Gradient Boosting Machine}, \textit{Categorical Boosting}, \textit{Histogram-Based Gradient Boosting Classifier}, and \textit{Adaptive Boosting Classifier}.

    \item \textbf{Bagging and Boosting Wrapper:} This group includes \textit{Naive Bayes Bagging}, \textit{Logistic Regression Bagging}, \textit{Support Vector Machine Bagging}, \textit{Decision Tree Bagging}, \textit{Multilayer Perceptron Bagging}, \textit{Naive Bayes Boosting}, \textit{Logistic Regression Boosting}, \textit{Support Vector Machine Boosting}, \textit{Decision Tree Boosting}, and \textit{Multilayer Perceptron Boosting}.

    \item \textbf{Imbalance-Aware Ensemble:} This group includes \textit{Balanced Random Forest Classifier}, \textit{Easy Ensemble Classifier}, \textit{Random Under Sampling Boosting Classifier}, and \textit{Balanced Bagging Classifier}.
\end{compactitem}

To reduce redundancy over the candidate classifiers, we used \textit{Classifier Output Difference} (COD), which measures behavioral distance by prediction disagreement~\cite{kuncheva2003measures,bowes2018software}.
For two classifiers $f_a$ and $f_b$ on the same test set of $n$ instances, COD is the proportion of instances on which their predicted labels differ:
$\text{COD}(f_a, f_b) = \frac{1}{n}\sum_{i=1}^{n}\mathbf{1}[f_a(x_i) \neq f_b(x_i)]$.
A COD score of 0 indicates identical predictions on all test instances, whereas a score approaching 1 indicates disagreement on nearly every prediction.

To compute COD distances that reflect the actual SDP behavior, we followed the methodology of Wan et al.~\cite{wan2024data}. 
We evaluated each candidate classifier on the 36 benchmark datasets using $5 \times 5$ repeated cross-validation. 
For each sample in every dataset, we obtained the majority-vote prediction across all CV folds from each classifier. 
We then computed pairwise COD values per dataset and averaged the resulting COD matrices across the 36 datasets to produce a single $40 \times 40$ distance matrix. 
This procedure ensured that the behavioral distances captured classifier disagreement patterns across the benchmark defect prediction tasks.

\begin{figure}[!htbp]
\centering
\includegraphics[width=\linewidth]{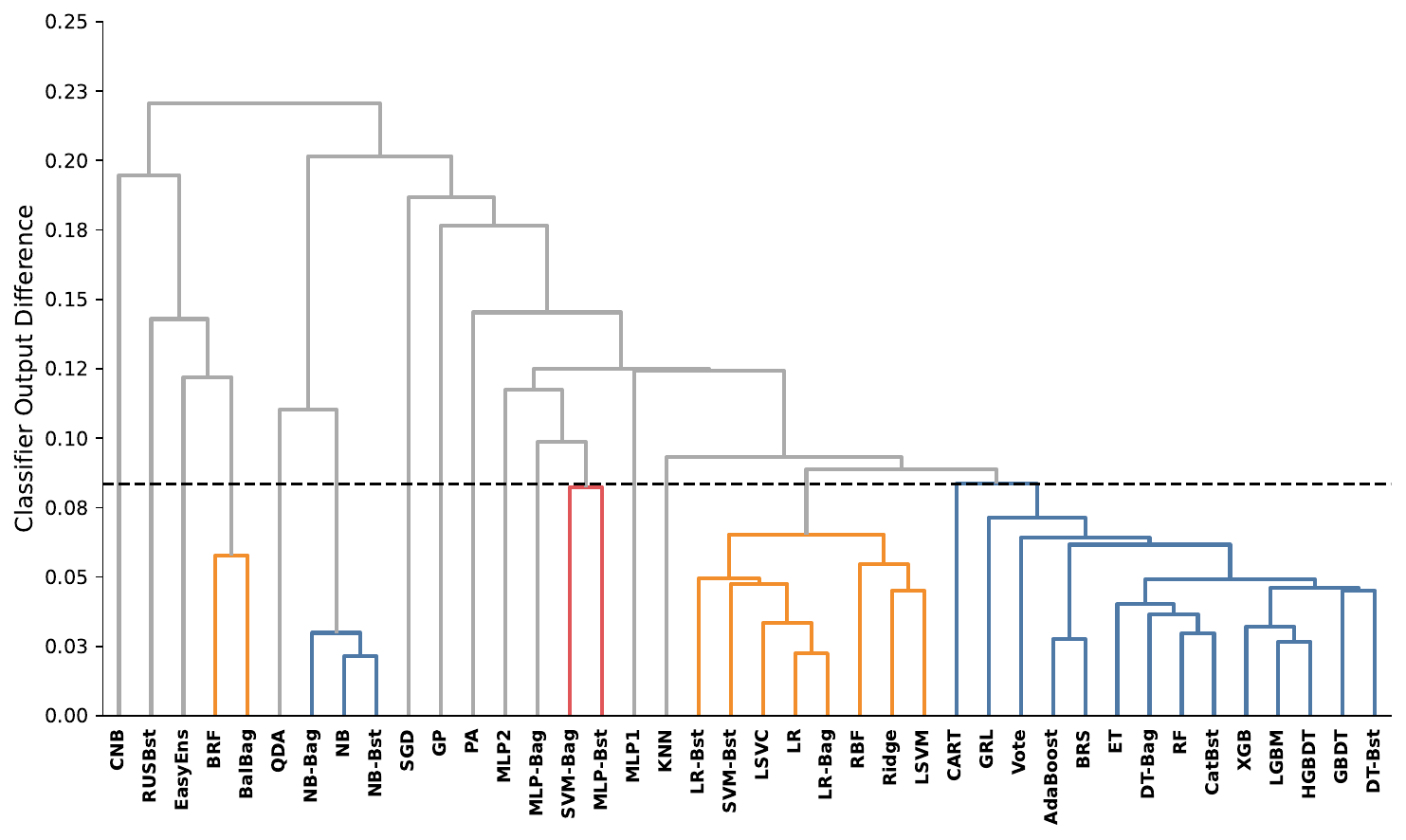}
\caption{Dendrogram of 40 candidate classifiers clustered using COD scores. Each leaf node at the bottom represents one candidate classifier. The vertical axis indicates COD distance: classifiers connected at shorter distances exhibit more similar prediction behavior. The dashed horizontal line marks the threshold used to partition the dendrogram. Each connected subtree below this threshold represents one cluster, and the leaves within the same subtree belong to the same cluster. This threshold yields 16 clusters, and one representative classifier remains in each cluster, resulting in the final 16 classifiers.}
\Description{Hierarchical clustering dendrogram of 40 candidate classifiers using classifier output difference. A dashed horizontal threshold indicates the level at which the tree is partitioned into 16 connected subtrees. Each subtree represents one cluster, and one classifier remains in each cluster.}
\label{fig:dendrogram}
\end{figure}

Using this pairwise COD distance matrix, we applied agglomerative hierarchical clustering with average linkage to group the 40 classifiers.
The clustering result is visualized as a dendrogram (Figure~\ref{fig:dendrogram}).
Each leaf node represents one candidate classifier.
The vertical axis shows the COD distance at which two branches merge; shorter merge distances indicate higher behavioral similarity.
We chose the COD cut point according to two criteria: preserving behavioral diversity among classifiers and avoiding over-representation of behaviorally redundant variants.
Specifically, we inspected the hierarchical COD dendrogram and selected the cut before a large increase in merge distance, where further merging would combine classifiers with visibly different prediction behaviors.
This cut produced 16 clusters.
The resulting clusters retained the major probability-generation mechanisms in the candidate pool, including probabilistic models, margin-based models, tree/rule-based models, neural networks, general ensembles, and imbalance-aware ensembles, while collapsing many wrapper variants that exhibited similar prediction behavior.
Thus, the number 16 was not chosen as a target model count, but resulted from a COD-based redundancy reduction criterion fixed before the RQ analyses.
Within each cluster, we selected one representative classifier using the following priority rules: preserving underrepresented learning categories, preferring base learners over wrappers when applicable, selecting the cluster medoid, and finally using the predefined candidate order to break ties.
As a result, we retained 16 representative classifiers from the original pool of 40 candidates.
Table~\ref{tab:model_categories} summarizes these representative classifiers.

\begin{table}[!htbp]
\centering
\small
\caption{Categorization of the 16 representative machine learning classifiers obtained by COD clustering.}
\label{tab:model_categories}
\renewcommand{\arraystretch}{1.4}
\begin{tabularx}{\textwidth}{@{}p{3.4cm} X@{}}
\toprule
\textbf{Category} & \textbf{Selected Representative Classifiers} \\
\midrule
Probabilistic Model
& Naive Bayes, Complement Naive Bayes, Gaussian Process Classifier, Quadratic Discriminant Analysis \\

Linear Model
& Logistic Regression, Stochastic Gradient Descent Classifier, Passive Aggressive Classifier \\

Lazy Learning
& KNN \\

Decision Tree
& Classification and Regression Tree \\

Neural Network
& One Hidden Layer Multilayer Perceptron, Two Hidden Layer Multilayer Perceptron, Multilayer Perceptron Bagging \\

SVM-Based
& Support Vector Machine Bagging \\

Ensemble
& Balanced Random Forest Classifier, Easy Ensemble Classifier, Random Under Sampling Boosting Classifier \\
\bottomrule
\end{tabularx}
\end{table}

Probabilistic classifiers estimate class likelihoods or decision boundaries from distributional assumptions.
Linear classifiers produce scores via a linear decision boundary transformed by a sigmoid link.
Lazy learning derives probabilities from local neighborhood density; KNN is the sole representative because it is the only lazy learner in the candidate pool and its distance-based probability mechanism is distinct from all other categories.
Decision trees estimate probabilities from leaf node class frequencies.
Neural Network classifiers output softmax probabilities from nonlinear hidden representations.
MLP Bagging is placed in this category because its base learner is an MLP that produces probabilities through the softmax mechanism.
The bagging wrapper reduces variance through bootstrap aggregation~\cite{breiman1996bagging} but does not alter the underlying probability generation process.
SVM-Based classifiers rely on margin transformations to convert signed margins into probabilities, a mechanism distinct from softmax, leaf frequency estimates, and likelihood-based estimates.
The Ensemble category uses class-balancing mechanisms that alter the effective class prior seen by the base learner, making its probability behavior different from that of the other six categories.

\subsubsection{Metrics}
\label{sec:metrics}

\begin{compactitem}
    \item \textbf{Uncertainty Metrics~\cite{wu2025uncertaintyzoo}:} Uncertainty reflects the confidence in a classifier's output.
    We used five common uncertainty metrics (Maximum Probability~\cite{hendrycks2016baseline}, Least Confidence~\cite{lewis1995sequential}, Margin Score~\cite{scheffer2001active}, Predictive Entropy~\cite{shannon1948mathematical}, and DeepGini~\cite{feng2020deepgini}), as summarized in Table~\ref{tab:uncertainty_metrics}.

    \item \textbf{Performance Metrics:} The study included MCC, Precision, Recall, FPR, AUC, and F1 score, which are widely adopted in SDP research~\cite{li2018progress,agrawal2018better,ozturk2017type}.

    \item \textbf{Calibration Metrics:} 
    Calibration measures the degree to which a model's prediction accurately reflects the true likelihood of an event.
    We used Negative Log Likelihood (NLL), Brier Score, and Expected Calibration Error (ECE)~\cite{minderer2021revisiting,gneiting2007strictly,glenn1950verification}.

    \item \textbf{Correlation Metrics:} We used three correlation coefficients: Pearson's $r$~\cite{schober2018correlation,de2016comparing}, Spearman's $\rho$~\cite{spearman1961proof}, and Kendall's $\tau$~\cite{kendall1938new}.
    
Table~\ref{tab:performance_calibration_metrics} provides descriptions for all performance, calibration, and correlation metrics.
\end{compactitem}

\begin{table}[!htbp]
\centering
\small
\caption{Uncertainty metrics in the study.}
\label{tab:uncertainty_metrics}
\renewcommand{\arraystretch}{1.16}
\setlength{\tabcolsep}{4pt}      

\begin{tabularx}{\textwidth}{@{}>{\raggedright\arraybackslash}p{2.6cm} >{\raggedright\arraybackslash}p{5.0cm} >{\raggedright\arraybackslash}X@{}}
\toprule
\textbf{Metric} & \textbf{Key Formulation} & \textbf{Description} \\ \midrule

\textbf{Maximum Probability} & 
$MP(x) = \max_{c} p(c|x)$ & 
This metric uses the highest predicted class probability as a certainty score. A lower value indicates higher uncertainty. \\

\rowcolor{lightgray}\textbf{Least Confidence} & 
$LC(x) = 1 - \max_{c} p(c|x)$ & 
This metric takes the complement of the most likely class probability. A higher value indicates higher uncertainty. \\

\textbf{Margin Score} & 
$M(x) = p(\hat{y}_1|x) - p(\hat{y}_2|x)$ & 
This metric computes the difference between the largest and second-largest predicted probabilities. A lower value indicates higher uncertainty. \\

\rowcolor{lightgray}\textbf{Predictive Entropy} & 
$H(y|x) = -\sum_{c=1}^{C} p(y_c|x) \log p(y_c|x)$ & 
This metric measures the dispersion of the predicted class distribution. A higher value indicates higher uncertainty. \\

\textbf{DeepGini} & 
$DG(x) = 1 - \sum_{c=1}^{C} p(y_c|x)^2$ & 
This metric is derived from the Gini impurity index. A higher value indicates a less concentrated predictive distribution and therefore higher uncertainty. \\ \bottomrule

\end{tabularx}
\end{table}

\begin{table}[!htbp]
\centering
\small
\caption{Performance, calibration, and correlation metrics in the study.}
\label{tab:performance_calibration_metrics}
\renewcommand{\arraystretch}{1.12}
\setlength{\tabcolsep}{3.5pt}      

\begin{tabularx}{\textwidth}{@{}>{\raggedright\arraybackslash}p{2.6cm} >{\raggedright\arraybackslash}p{5.0cm} >{\raggedright\arraybackslash}X@{}}
\toprule
\textbf{Metric} & \textbf{Formulation} & \textbf{Description} \\ \midrule

\multicolumn{3}{c}{\textsc{\textbf{Performance Metrics}}} \\ \midrule

\textbf{MCC} & 
\resizebox{4.6cm}{!}{$MCC = \frac{TP \cdot TN - FP \cdot FN}{\sqrt{(TP+FP)(TP+FN)(TN+FP)(TN+FN)}}$} & 
The Matthews Correlation Coefficient is a balanced measure that is less sensitive to class imbalance, ranging from -1 to +1. \\

\rowcolor{lightgray}\textbf{Precision} & 
$P = \frac{TP}{TP + FP}$ & 
This metric reports the ratio of correctly predicted positives to all predicted positives. \\

\textbf{Recall} & 
$R = \frac{TP}{TP + FN}$ & 
This metric, also known as Probability of Detection, reports the ratio of correctly predicted positives to all actual positives. \\

\rowcolor{lightgray}\textbf{F1 score} & 
$F1 = 2 \cdot \frac{Precision \cdot Recall}{Precision + Recall}$ & 
The F1 score is the harmonic mean of Precision and Recall, balancing both false positives and false negatives. \\

\textbf{FPR} & 
$FPR = \frac{FP}{FP + TN}$ & 
This metric measures the ratio of actual negatives incorrectly classified as positive. A larger FPR indicates more false alarms during inspection. \\

\rowcolor{lightgray}\textbf{AUC} & 
$\int_{0}^{1} TPR(t) \, d(FPR(t))$ & 
The Area Under the ROC Curve (AUC) evaluates the classifier's ability to discriminate between classes across all thresholds.
\\ 
\midrule

\multicolumn{3}{c}{\textsc{\textbf{Calibration Metrics}}} \\ \midrule

\textbf{ECE} & 
$\sum_{m=1}^{M} \frac{|B_m|}{n} \left| \bar{y}(B_m) - \bar{p}(B_m) \right|$ & 
The Expected Calibration Error measures the gap between the empirical positive rate and the mean positive-class probability across $M$ bins. \\

\rowcolor{lightgray}\textbf{NLL} & 
$NLL = -\frac{1}{n}\sum_{i=1}^{n}\log p(y_i \mid x_i)$ & 
The Negative Log Likelihood evaluates probabilistic prediction error. Lower values indicate better probabilistic predictions. \\

\textbf{Brier Score} & 
$BS = \frac{1}{n}\sum_{i=1}^{n}\bigl(p(y{=}1 \mid x_i) - y_i\bigr)^2$ & 
The Brier Score measures the mean squared error between predicted probabilities and observed outcomes. Lower values indicate better probabilistic predictions. \\ \midrule

\multicolumn{3}{c}{\textsc{\textbf{Correlation Coefficients}}} \\ \midrule

\textbf{Spearman's $\rho$} &
$1 - \frac{6 \sum d_i^2}{n(n^2 - 1)}$ &
This nonparametric coefficient measures monotonic correspondence between uncertainty and performance. \\

\rowcolor{lightgray}\textbf{Kendall's $\tau$} &
$\frac{C - D}{\sqrt{(C+D+T_u)(C+D+T_v)}}$ &
This coefficient measures ordinal correlation via concordant/discordant pairs and is less sensitive than Spearman's $\rho$ to extreme values. \\

\textbf{Pearson's $r$} &
$\frac{\sum(x_i - \bar{x})(y_i - \bar{y})}{\sqrt{\sum(x_i - \bar{x})^2 \sum(y_i - \bar{y})^2}}$ &
This coefficient measures the linear correlation between variables and serves as a baseline for linear dependency. \\ \bottomrule

\end{tabularx}
\end{table}

\subsubsection{Experimental Protocol}
\label{sec:experimental_protocol}
We repeated all experiments with 30 independent random seeds and reported the average over these runs.
For classifiers that returned only a decision function, we transformed the raw score $s$ into a score vector with two entries, $[1-\sigma(s), \sigma(s)]$, where $\sigma(s)=1/(1+\exp(-s))$.
This choice is motivated by Platt's sigmoid calibration method for SVM decision values, which uses a sigmoid function to map margin scores to probability estimates~\cite{platt1999probabilistic,niculescu2005predicting,wu2004probability,zadrozny2002transforming}.
Here, a decision function denotes an uncalibrated score, such as a signed margin or distance from the decision boundary, rather than a probability.
We did not assume that either the directly reported probabilities or the sigmoid-transformed scores were calibrated; our calibration analysis evaluated probability calibration separately.

For each classifier and dataset, we recorded three groups of quantities.
First, we computed the six predictive performance metrics listed in Table~\ref{tab:performance_calibration_metrics}.
Second, we computed the three calibration metrics from the predicted probabilities.
Third, we computed the five UQ metrics.
After seed averaging, each WPDP classifier and dataset pair contributed one observation containing performance, calibration, and UQ values.

For RQ1, we correlated each UQ metric with each performance metric using Pearson's $r$, Spearman's $\rho$, and Kendall's $\tau$.
The global analysis used all classifier and dataset observations.
Analyses by dataset collection computed the same correlations separately for AEEEM, NASA, PROMISE, and ReLink.
Granularity analyses followed the dataset mapping described in Section~\ref{sec:study_design}: AEEEM and PROMISE are class-level datasets, NASA is function- or method-level, and ReLink is file-level.
For RQ2, we computed Spearman correlations between the performance metrics and the calibration metrics across the same WPDP observations.

For RQ3, we evaluated CPDP without transfer learning~\cite{nam2013transfer}.
Within each included dataset collection, one dataset served as the source project and a different dataset from the same dataset collection served as the target project.
We treated source-to-target pairs as ordered, so we evaluated both $A \rightarrow B$ and $B \rightarrow A$ for two datasets $A$ and $B$.
For each ordered pair, we trained each of the 16 classifiers on the source project and evaluated it directly on the target project.
We aligned source and target features by common feature names, fitted scaling parameters only on the source project, and then applied them to the target project.
The CPDP analysis included AEEEM, NASA, and PROMISE, covering 32 datasets.
We excluded ReLink from CPDP because its four datasets formed two feature-incompatible subgroups.
Because these two subgroups shared no common features, ReLink did not provide enough source-to-target pairs within the same dataset collection for a stable CPDP analysis.
\subsection{RQ1: \rqone}
\label{sec:rq1}

RQ1 and RQ2 are conducted under WPDP, where the training and test data come from the same project; RQ3 then revisits these relationships under CPDP.
To address RQ1, we examined whether uncertainty estimates were correlated with predictive performance under WPDP and how this correlation varied across classifier categories, granularities, and datasets.
We organized the evaluation around the following three dimensions.

\begin{compactitem}
    \item \textbf{RQ1.1 (Global Correlation).} 
    We first computed global correlations between UQ and performance across all classifier and dataset pairs.
    \item \textbf{RQ1.2 (Category-Level Alignment).} We assessed whether alignment between UQ and performance differed across classifier categories.
    \item \textbf{RQ1.3 (Granularity-Level Correlation).} We examined whether correlations between UQ and performance differed across dataset collections (AEEEM, NASA, PROMISE, ReLink) and across prediction granularities (class-level, function/method-level, file-level). 
\end{compactitem}

\subsubsection{RQ1.1: Global Correlation Between UQ and Performance}
\label{sec:rq1_1}

\begin{figure}[!htbp]
    \centering
    \includegraphics[width=\linewidth, height=0.84\textheight, keepaspectratio]{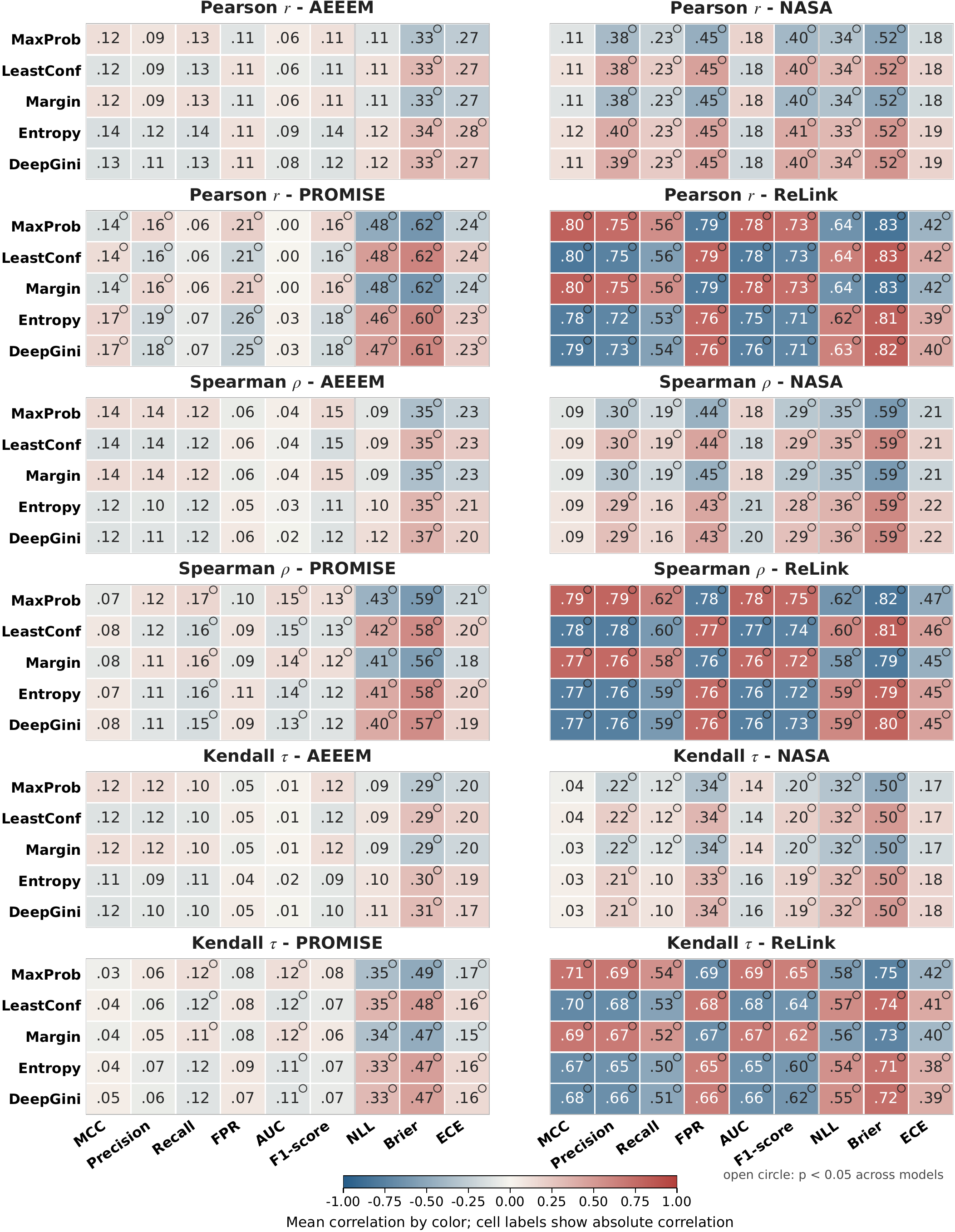}
    \caption{\textbf{Correlations between UQ and performance by dataset collection and correlation method.} Each heatmap shows the mean correlation across the classifiers between five UQ metrics (rows) and nine target metrics (columns). Color scale: blue ($-1$) to red ($+1$). Cell labels report absolute correlation magnitudes on the original correlation scale; the sign is encoded by color. Open circles mark cells where the 16 classifier-level correlations pass a one-sample $t$ test at $p < 0.05$, indicating that the mean correlation across classifiers is statistically distinguishable from zero. ReLink shows larger correlations than the other three collections. 
    }
    \Description{Twelve heatmaps arranged in an optimized 6x2 grid on a portrait page, showing correlations between five UQ metrics and nine target metrics for AEEEM, NASA, PROMISE, and ReLink across Pearson, Spearman, and Kendall methods. Cell labels show absolute correlation magnitudes on the original correlation scale; the heatmap color encodes the sign.}
    \label{fig:heatmap_all}
\end{figure}

The experiment used 36 datasets and 16 classifiers.
We computed Pearson's $r$, Spearman's $\rho$, and Kendall's $\tau$ between the five UQ metrics and six performance metrics.
Figure~\ref{fig:heatmap_all} shows the correlation matrices for Pearson $r$, Spearman $\rho$, and Kendall $\tau$ across each dataset collection. 
The heatmap color scale represents the average correlation magnitude and polarity across the 16 classifiers. 
Cell labels report absolute correlation magnitudes on the original correlation scale, with the sign encoded by color.
Open circles denote cells where the classifier-level correlations achieve statistical significance via a one-sample $t$ test ($p < 0.05$).
Aggregated across all dataset collections, the majority of correlations between UQ and performance remained marginal.

The most prominent signal involved the false positive rate (FPR). 
Maximum Probability had a negative correlation with FPR (Spearman $\rho \approx -0.48$; Kendall $\tau \approx -0.38$), whereas uncertainty-oriented metrics, such as Predictive Entropy and Least Confidence, correlated positively with FPR ($\rho \approx +0.47$ and $+0.48$, respectively). 
Given that higher Maximum Probability denotes lower uncertainty, this pattern was directionally consistent: classifiers with greater predictive uncertainty on a dataset tended to yield higher false positive rates. 
Similarly, AUC correlated negatively with Predictive Entropy (Spearman $\rho \approx -0.44$; Kendall $\tau \approx -0.32$), suggesting that more dispersed predictive distributions were associated with weaker discrimination. 
Conversely, Maximum Probability correlated positively with AUC ($\rho \approx +0.44$).

In contrast, the correlation with MCC was marginal and inconsistent across coefficients (Spearman $\rho \approx +0.13$; Pearson $r \approx +0.23$). 
Because MCC is a balanced metric reflecting the full confusion matrix, its weak correlation with mean UQ suggests that aggregate uncertainty is a weak proxy for balanced classification effectiveness. 
Precision, Recall, and F1 score likewise had negligible global correlations; notably, for the F1 score, all Spearman correlations with the five UQ metrics were near zero ($\lvert\rho\rvert \approx 0$).

\begin{findingbox}
\textbf{Finding 1 (WPDP):} 
Globally, correlations between UQ and performance in WPDP were generally weak, appearing consistent only for specific metric pairs. 
These results demonstrate that UQ is \emph{not} a universal proxy for classifier performance in SDP; rather, its utility is highly contingent upon the choice of evaluation metric and the specific dataset context.
\end{findingbox}

\subsubsection{RQ1.2: Category-Level Alignment Between UQ and Performance}
\label{sec:rq1_2}

While Finding~1 showed that aggregate correlations between UQ and performance in WPDP were generally marginal, we next investigated whether this trend was maintained across classifier categories. 
Identifying specific categories that yield performance-correlated UQ is essential for the practical application of uncertainty estimates in model selection or performance assessment in SDP.

\noindent\textbf{Experimental Design.}
We computed an alignment score between UQ and performance for each classifier and each performance metric (AUC, MCC, F1 score, Precision, Recall, FPR).
The score was the mean Spearman correlation between dataset-level UQ values and dataset-level performance metric values across the five UQ metrics. 

\begin{figure}[!tbp]
\centering
\setlength{\tabcolsep}{0pt}
\newcommand{\rw}{0.49\linewidth}
\begin{tabular}{@{}cc@{}}
\includegraphics[width=\rw]{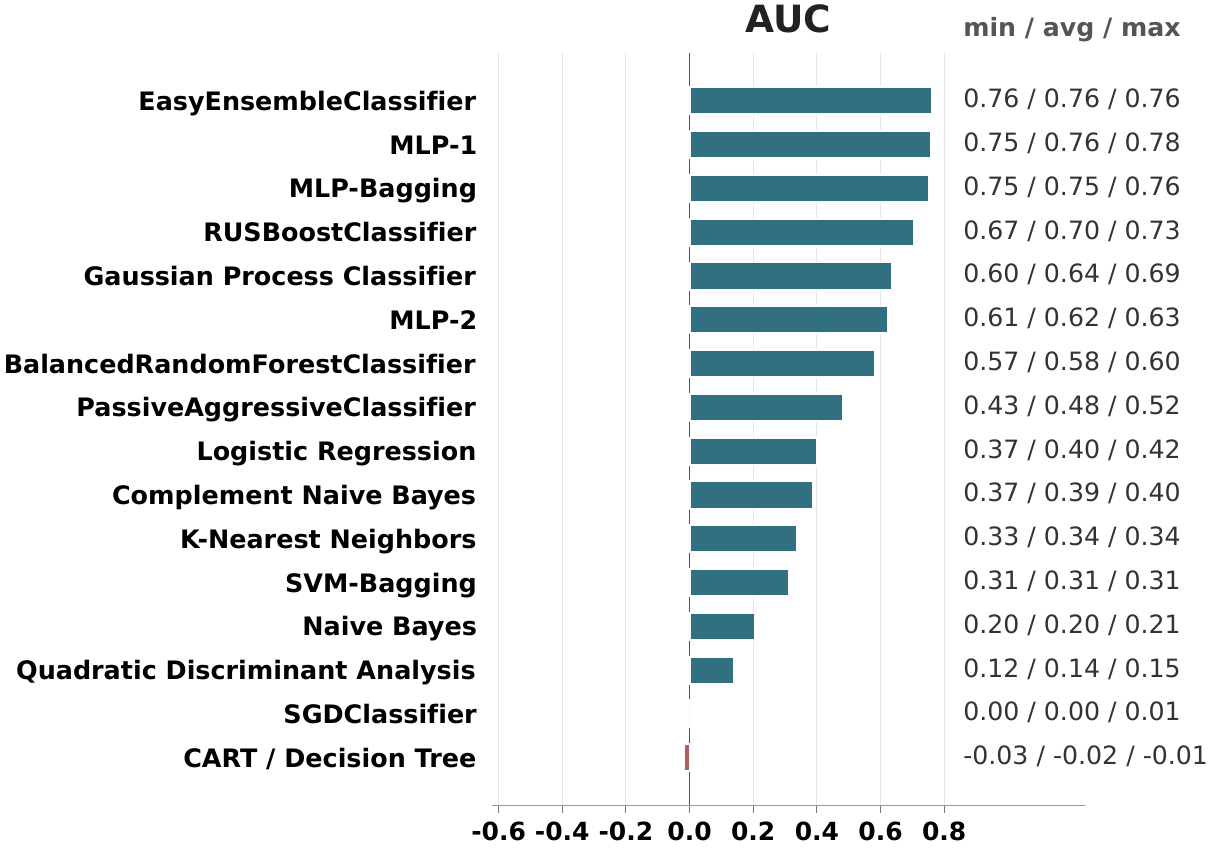} &
\includegraphics[width=\rw]{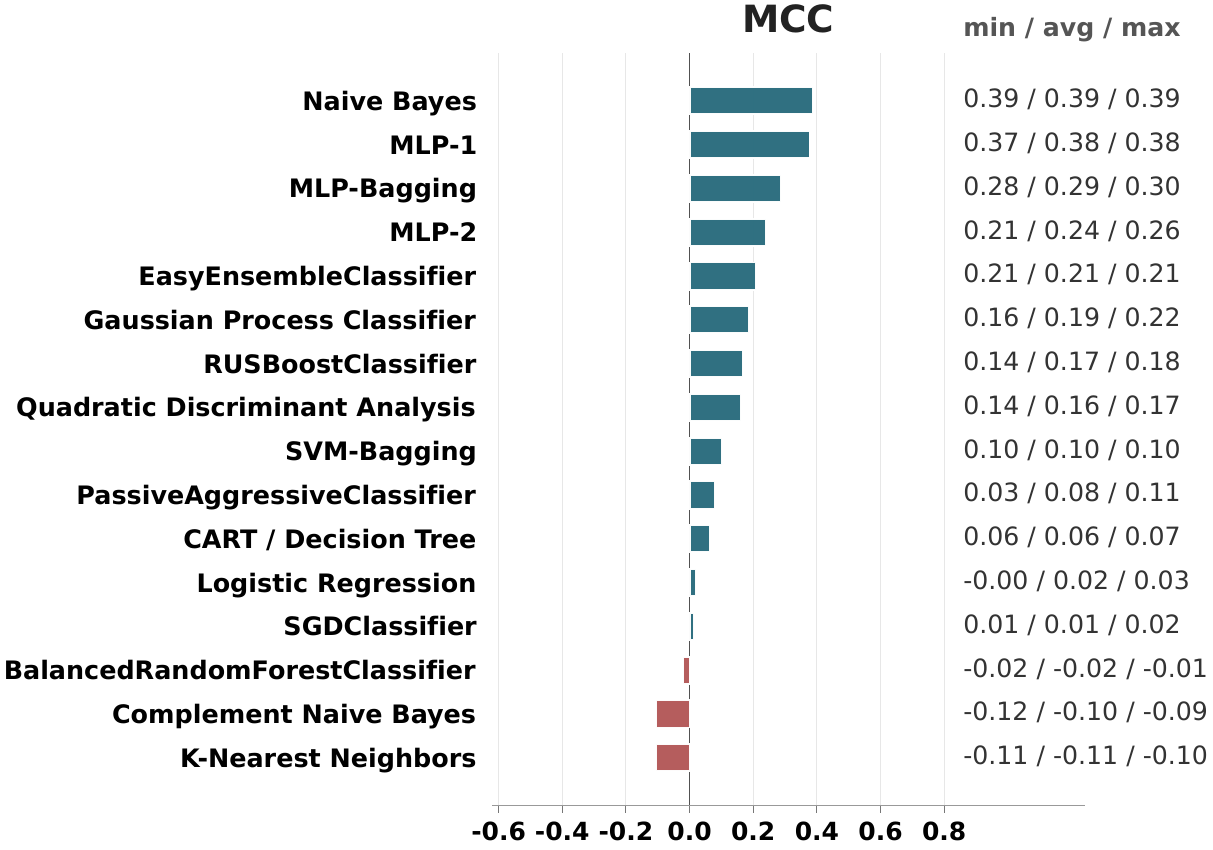} \\
\textit{(a) AUC} & \textit{(b) MCC} \\[0.08cm]
\includegraphics[width=\rw]{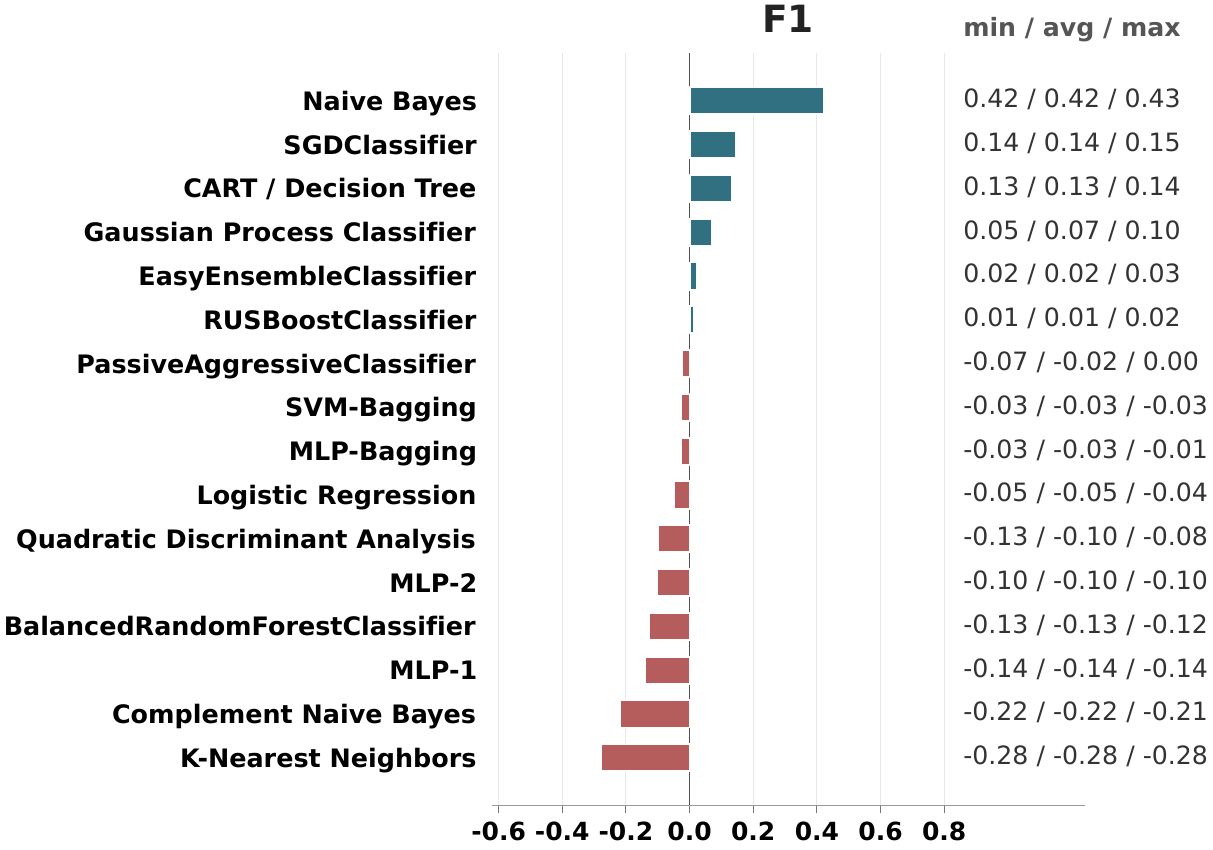} &
\includegraphics[width=\rw]{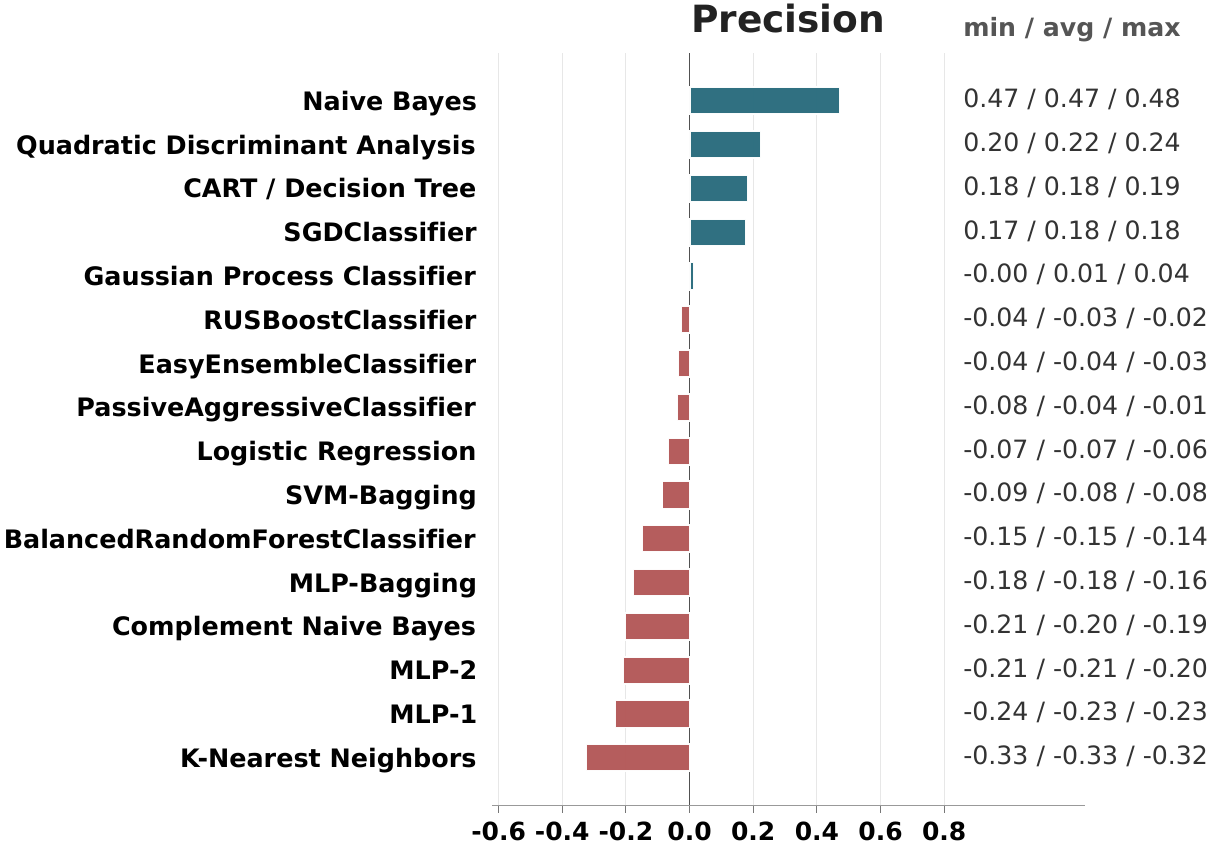} \\
\textit{(c) F1 score} & \textit{(d) Precision} \\[0.08cm]
\includegraphics[width=\rw]{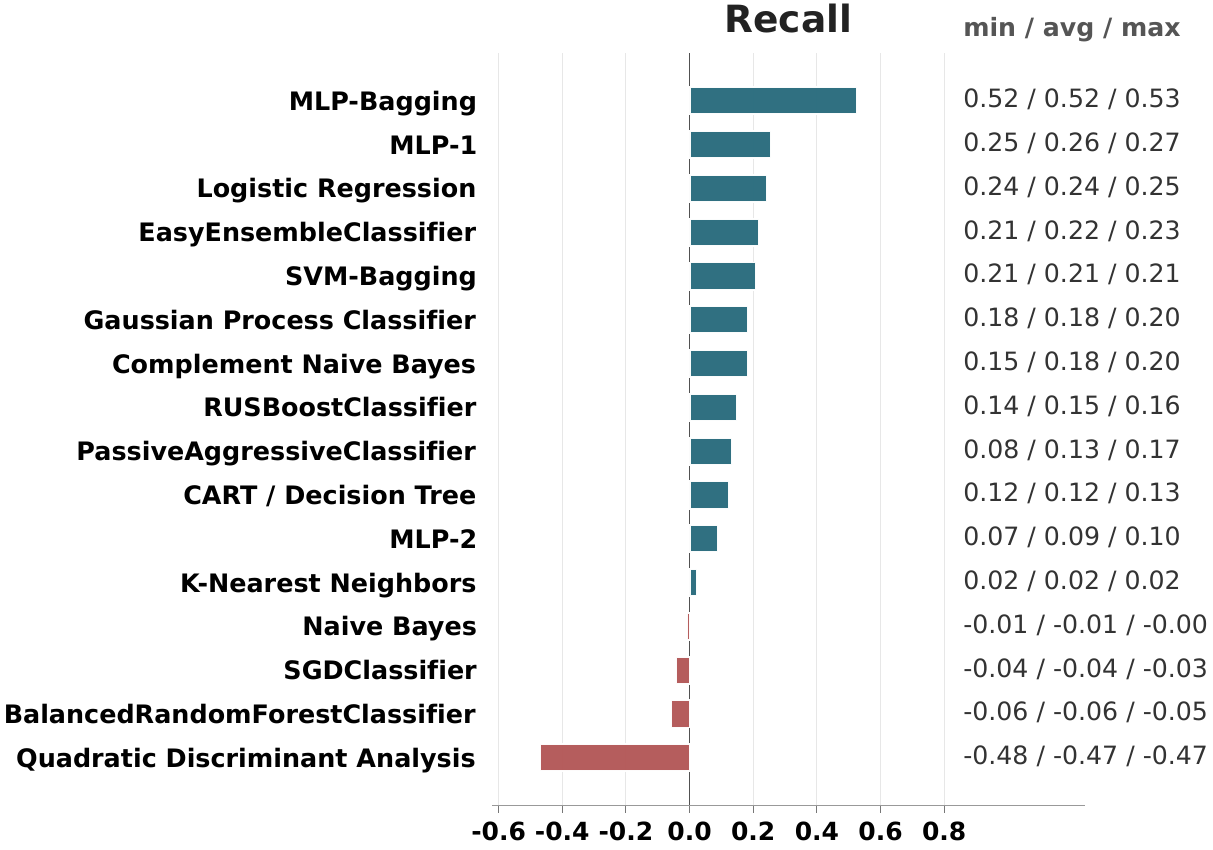} &
\includegraphics[width=\rw]{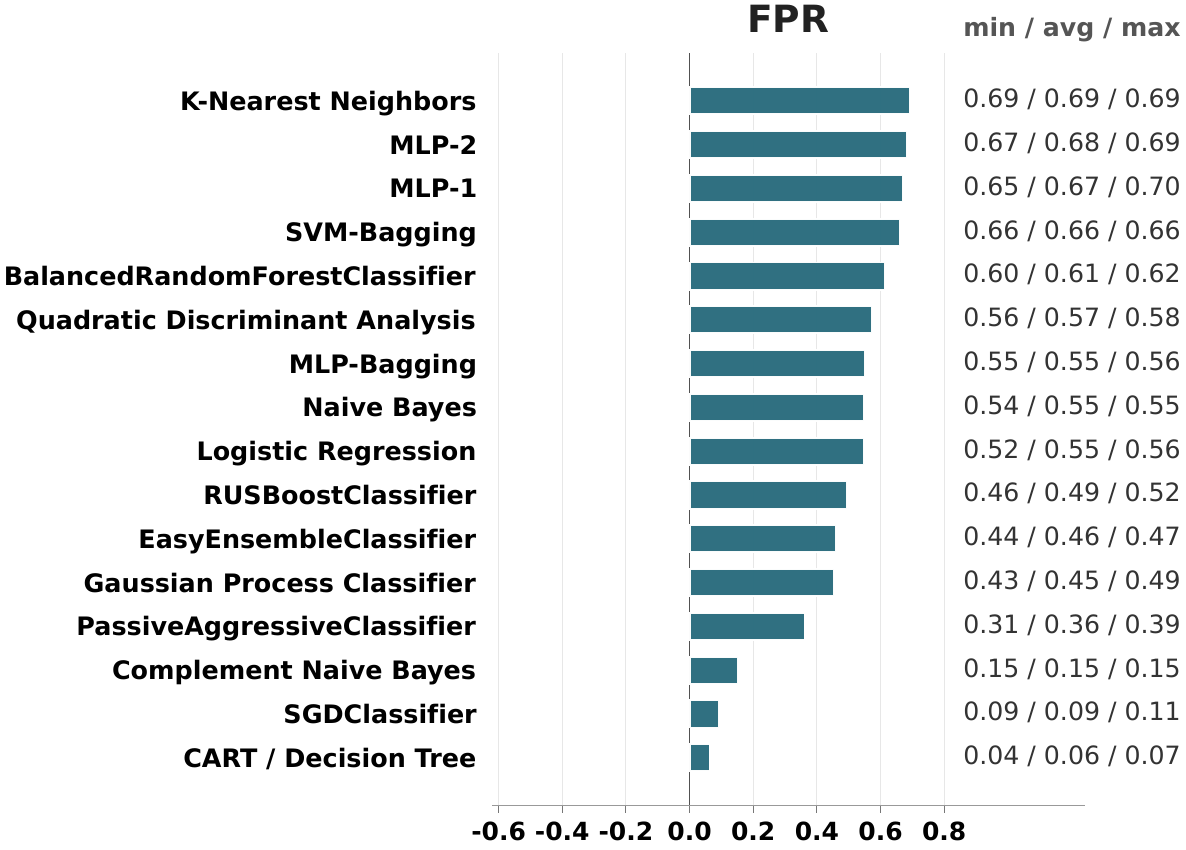} \\
\textit{(e) Recall} & \textit{(f) FPR} \\
\end{tabular}
\vspace{-0.05cm}
\caption{\textbf{Alignment scores between UQ and performance across six performance metrics.} Each panel shows the alignment score (Spearman correlation between dataset-level UQ values and performance metric values, sign-adjusted so that higher is better) with the minimum, average, and maximum across the five UQ metrics displayed for each classifier. Classifiers are sorted by the average alignment score within each panel.}
\Description{Six horizontal bar charts arranged in a 3x2 grid, showing alignment scores between UQ and performance for AUC, MCC, F1 score, Precision, Recall, and FPR. Ensemble methods occupy top positions in most panels; Neural Network and probabilistic classifiers cluster at the bottom.}
\label{fig:uq_alignment_all}
\end{figure}

\noindent\textbf{Result.}
Figure~\ref{fig:uq_alignment_all} illustrates the alignment scores between UQ and performance across the six evaluation metrics. We observed four key patterns. 
First, AUC alignment was positive: most classifiers showed moderate-to-strong alignment between UQ and AUC, led by neural networks and imbalance-aware ensembles (MLP~1 and EasyEnsemble: $0.76$; MLP-Bagging: $0.75$; RUSBoost: $0.70$; Gaussian Process: $0.64$), while only CART ($-0.02$) and SGDClassifier ($0.00$) were near zero.
Second, FPR alignment was also positive and often strong (e.g., KNN $0.69$, MLP~2 $0.68$, MLP~1 $0.67$, SVM Bagging $0.66$), consistent with Finding~1, where UQ correlated most strongly with FPR and AUC.
Third, alignment with threshold-dependent metrics (MCC, Precision, F1 score) was weaker and sometimes negative even for classifiers with strong AUC alignment (e.g., MLP~1 Precision alignment $-0.23$, KNN Precision $-0.33$), so a strong UQ--AUC signal did not imply a useful signal for threshold-based decisions.
Fourth, the min--max range across the five UQ metrics was narrow for almost all classifiers ($\le 0.05$), indicating that alignment conclusions were robust to the choice of UQ metric; in binary classification the five probability-based UQ metrics are near-monotone transforms of one another.

\begin{compactitem}
    \item \textbf{Ensemble.}
The ensemble category contained three imbalance-aware classifiers (RUSBoostClassifier, EasyEnsembleClassifier, BalancedRandomForestClassifier) and showed among the highest AUC alignment in the benchmark.
EasyEnsembleClassifier had the largest AUC alignment ($0.76$), followed by RUSBoostClassifier ($0.70$) and BalancedRandomForestClassifier ($0.58$).
The min--max ranges were narrow ($\le 0.06$ for all three), confirming that their UQ signals were consistent across all five UQ metrics.
This strong AUC alignment did not, however, carry over to MCC: EasyEnsemble ($0.21$) and RUSBoost ($0.17$) led, while BalancedRandomForest was near zero ($-0.02$).
Thus, imbalance-aware ensembles provided a strong and consistent UQ signal for AUC but only a weak one for balanced classification performance.

    \item \textbf{Linear Model.}
PassiveAggressiveClassifier ($0.48$ for AUC) and Logistic Regression ($0.40$) showed moderate positive AUC alignment, while SGDClassifier was neutral ($0.00$).
For MCC, all three were weak (PassiveAggressiveClassifier $0.08$, Logistic Regression $0.02$, SGDClassifier $0.01$).
The linear category therefore aligned with AUC but not with balanced classification performance.
Differences in score generation and probability transformation affected how uncertainty metrics correlated with performance.

    \item \textbf{Decision Tree.}
CART had near-zero AUC alignment ($-0.02$) and weak alignment for the other metrics (MCC $0.06$, Precision $0.18$, FPR $0.06$).
It produced the most concentrated probabilities of any classifier (mean Maximum Probability $0.980$, Predictive Entropy $0.048$).
This concentration left almost no dynamic range for UQ metrics, which reduced their capacity to correlate with any downstream performance metric.
Its min--max range across the five UQ metrics was narrow (AUC: $-0.03$ to $-0.01$), so the near-zero alignment held across all five UQ metrics.
CART was thus the clearest case of an uninformative UQ signal in the benchmark.

    \item \textbf{Lazy Learning.}
KNN showed moderate AUC alignment ($0.34$) and strong FPR alignment ($0.69$), but negative alignment for Precision ($-0.33$) and MCC ($-0.11$).
Its distance-based probability estimates therefore produced UQ signals that tracked ranking and false-positive behavior but were inversely related to threshold-dependent precision.

    \item \textbf{Probabilistic Model.}
Gaussian Process Classifier ($0.64$ for AUC) showed the highest AUC alignment in this category, followed by Complement Naive Bayes ($0.39$), Naive Bayes ($0.20$), and QDA ($0.14$).
Naive Bayes was unusual in aligning more strongly with MCC ($0.39$) and Precision ($0.47$) than with AUC, the opposite of most classifiers, whereas Complement Naive Bayes showed negative MCC ($-0.10$) and Precision ($-0.20$) alignment.
These classifiers produced concentrated probabilities under distributional assumptions that are violated by software metrics (Naive Bayes mean Maximum Probability $0.979$, QDA $0.968$).
The present analysis does not isolate whether the Gaussian Process pattern came from the kernel model, the probability transformation, or averaging across random seeds.

    \item \textbf{Neural Network.}
All three neural classifiers showed strong AUC alignment (MLP~1: $0.76$, MLP-Bagging: $0.75$, MLP~2: $0.62$) and among the highest MCC alignment in the benchmark (MLP~1: $0.38$, MLP-Bagging: $0.29$, MLP~2: $0.24$), with only Naive Bayes ($0.39$) ranking marginally higher on MCC.
In contrast to the imbalance-aware ensembles, the neural classifiers thus produced UQ signals that aligned with both ranking (AUC) and balanced (MCC) performance.
The three architectures behaved consistently, indicating that the softmax probability mechanism yielded informative uncertainty signals regardless of depth or bagging.

    \item \textbf{SVM-Based.}
SVM Bagging showed moderate AUC alignment ($0.31$), strong FPR alignment ($0.66$), but near-zero MCC ($0.10$) and Precision ($-0.08$) alignment.
Despite its concentrated probability estimates (mean Maximum Probability $0.945$), its margin-derived UQ signal still tracked ranking and false-positive behavior, though not threshold-dependent performance.
\end{compactitem}

\begin{findingbox}
\textbf{Finding 2 (WPDP):} Alignment between UQ and performance varied more by target metric than by classifier category under WPDP.
For AUC, most classifiers showed moderate-to-strong positive alignment, led by neural networks (MLP~1: $0.76$, MLP-Bagging: $0.75$) and imbalance-aware ensembles (EasyEnsemble: $0.76$, RUSBoost: $0.70$); only CART ($-0.02$) and SGDClassifier ($0.00$) were near zero.
The same classifiers aligned more weakly, and sometimes negatively, with threshold-dependent metrics such as MCC and Precision.
The alignment was consistent across the five UQ metrics for all classifiers (min--max ranges $\le 0.10$), reflecting that probability-based UQ metrics are near-monotone transforms of one another in binary classification.
In practice, UQ should be interpreted as a metric-specific signal---reliable for AUC and FPR but not for threshold-dependent metrics such as MCC and F1 score---rather than as a property of the uncertainty metric alone.
\end{findingbox}

\subsubsection{RQ1.3: Granularity-Level Correlation Between UQ and Performance}
\label{sec:rq1_3}

While Finding~2 characterized how alignment between UQ and performance varied across classifier categories and target metrics, we next investigated whether it was also influenced by prediction granularity.
SDP datasets typically use different granularities, such as class-level modules (AEEEM, PROMISE), function- or method-level modules (NASA), and file-level modules (ReLink). 
For UQ metrics to be effective in performance assessment, it is essential to determine whether their usefulness changes across prediction granularities. 

\noindent\textbf{Experimental Design.}
We grouped the 36 datasets into three granularities: class (AEEEM and PROMISE, 20 datasets), function (NASA, 12 datasets), and file (ReLink, 4 datasets).
The three groups contained 20, 12, and 4 datasets, respectively, and we evaluated each group with all 16 classifiers.
For each group, we computed Spearman correlations between each UQ metric and each performance metric per classifier across the datasets in that group.
We then reported the mean correlation across the 16 classifiers with a one-sample $t$ test for significance.

\begin{figure}[!tbp]
\centering
\includegraphics[width=\linewidth]{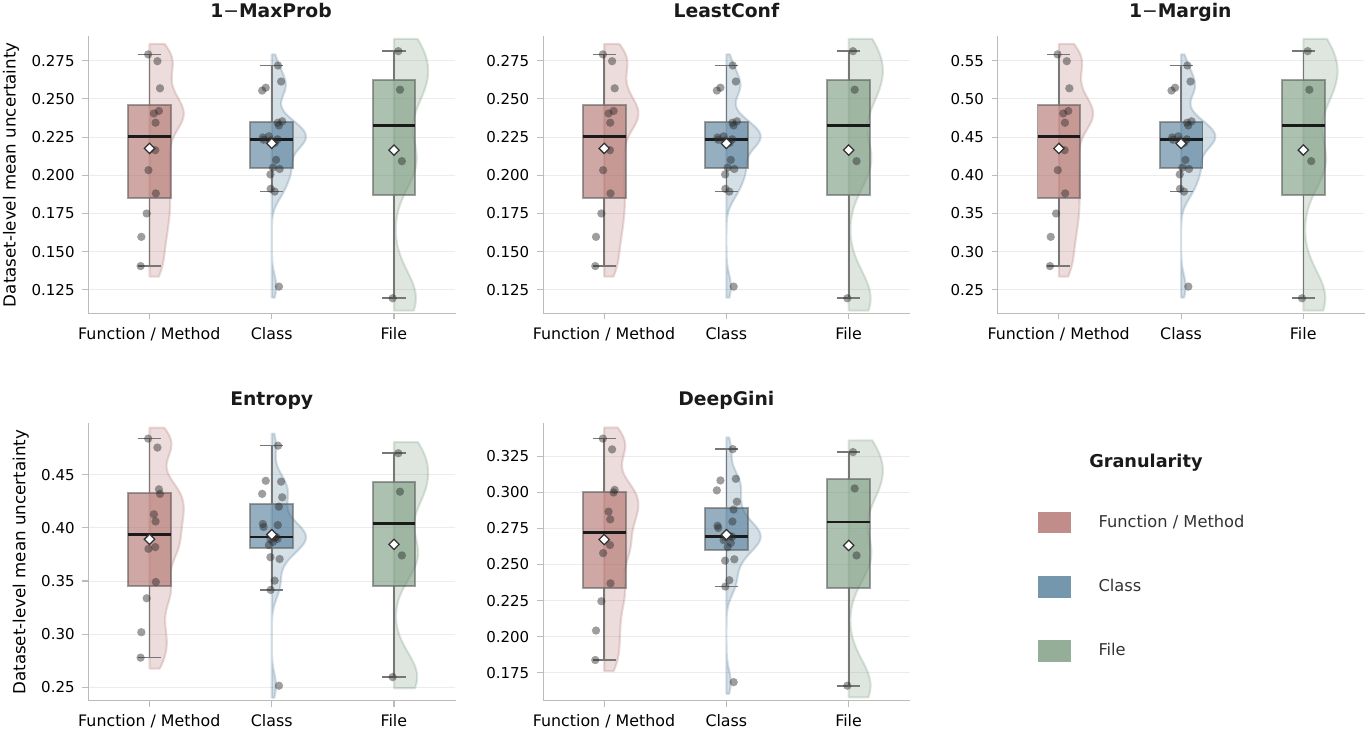}
\caption{\textbf{Uncertainty distribution by prediction granularity-aware dataset group.} Each panel shows the dataset-level mean of one UQ metric for class (AEEEM and PROMISE), function/method (NASA), and file (ReLink) datasets. Box plots show median and interquartile range; half violin plots show density; points are individual datasets; diamonds mark means. The file-level ReLink group has the widest uncertainty spread. 
}
\Description{Five panels arranged in a 2x3 grid showing box plots with half violin overlays for five UQ metrics across three granularity levels. The file-level group shows the widest spread in every panel.}
\label{fig:uq_granularity_distribution}
\end{figure}

\noindent\textbf{Result.}
Figure~\ref{fig:uq_granularity_distribution} illustrates the distribution of dataset-level mean uncertainty across the three granularity levels. 
While median uncertainty levels remained relatively consistent across groups, their variance differed substantially: the file-level ReLink group exhibited the widest dispersion, whereas the class-level and function/method-level groups demonstrated more concentrated distributions. 
This pattern suggests that the correlation between UQ and performance depends less on a uniform shift in confidence than on the degree of uncertainty variation across classifiers and datasets within a specific group.

\begin{compactitem}
    \item \textbf{Class Level.}
At the class level, disaggregating the two dataset collections revealed distinct profiles. AEEEM remained weak: MaxProb vs.~AUC was small and not significant ($\rho \approx 0.13$, n.s.), and among its UQ--performance correlations only FPR reached significance (MaxProb vs.~FPR $\rho \approx -0.33$, $p < 0.01$). PROMISE was stronger: both MaxProb vs.~AUC ($\rho \approx 0.33$, $p < 0.001$) and MaxProb vs.~FPR ($\rho \approx -0.28$, $p < 0.001$) were significant, while correlations with F1 score and MCC stayed weak.
Predictive Entropy mirrored these patterns with opposite signs (PROMISE Entropy vs.~AUC $\rho \approx -0.32$).
Mean class-level performance was moderate (MCC $\approx 0.30$, AUC $\approx 0.74$).
The divergence between AEEEM and PROMISE likely reflected differences in dataset composition: AEEEM comprises five projects from the Eclipse ecosystem with relatively homogeneous metric distributions, leaving limited cross-dataset variance for UQ to track, whereas PROMISE spans 15 projects with more diverse defect ratios and feature profiles, providing enough variability for a detectable UQ--performance correlation.

    \item \textbf{Function Level.}
At the function level (NASA), correlations were stronger than at the class level and split into two distinct patterns.
Maximum Probability correlated positively with AUC ($\rho \approx 0.49$, $p < 0.001$), but negatively with F1 score ($\rho \approx -0.33$, $p < 0.001$), Precision ($\rho \approx -0.36$, $p < 0.001$), and FPR ($\rho \approx -0.66$, $p < 0.001$).
The negative F1 score and Precision correlations indicated that higher-certainty classifiers tended to have worse threshold-dependent performance at the default cutoff; yet the large negative FPR correlation means higher confidence is correlated with fewer false positives.
Entropy-oriented metrics mirrored these patterns (e.g., Predictive Entropy vs.~FPR $\rho \approx 0.67$, $p < 0.001$).
Mean function/method-level performance was the lowest of the three groups (MCC $\approx 0.26$, AUC $\approx 0.74$).
The opposing signs between ranking-based (AUC) and threshold-dependent (F1 score, Precision) metrics indicated that at the function level, classifier confidence improved ranking ability but did not translate into better decision-boundary performance at the default cutoff.

    \item \textbf{File Level.}
At the file level (ReLink), correlations between UQ and performance were the largest in the benchmark.
Maximum Probability correlated positively with AUC ($\rho \approx 0.79$), MCC ($\rho \approx 0.78$), Precision ($\rho \approx 0.76$), F1 score ($\rho \approx 0.70$), and Recall ($\rho \approx 0.63$), and negatively with FPR ($\rho \approx -0.70$), all at $p < 0.001$.
Predictive Entropy mirrored this pattern (e.g., Entropy vs.~AUC $\rho \approx -0.75$, $p < 0.001$).
Mean file-level performance was the highest of the three groups (MCC $\approx 0.51$, AUC $\approx 0.80$).
The strong correlations at the file level may be attributed to two factors: the ReLink datasets exhibited the widest uncertainty dispersion among all groups (Figure~\ref{fig:uq_granularity_distribution}), providing sufficient dynamic range for UQ metrics to differentiate dataset difficulty, and the coarser file-level granularity may have amplified differences in defect characteristics across datasets.
This group contained only four datasets, so we treated the result as ReLink-specific evidence rather than as evidence that file-level prediction increases the correlation between UQ and performance.
\end{compactitem}
These granularity-level results indicated that the strength of UQ--performance correlation increased with the heterogeneity of the underlying data: AEEEM (homogeneous Eclipse projects) showed the weakest signals, PROMISE and NASA were progressively stronger, and ReLink (the widest uncertainty dispersion) showed the strongest.
Because granularity and dataset collection were coupled in this benchmark, these patterns should be interpreted as joint effects of both factors rather than as isolated granularity effects.

\begin{findingbox}
\textbf{Finding 3 (WPDP):} Correlations between UQ and performance differed across granularities and grew with dataset heterogeneity.
At the class level, AEEEM showed no statistically significant UQ vs.~AUC correlation ($\rho \approx 0.13$, n.s.), while PROMISE showed a significant positive correlation (MaxProb vs.~AUC $\rho \approx 0.33$, $p < 0.001$).
At the function level, MaxProb vs.~AUC was stronger ($\rho \approx 0.49$, $p < 0.001$), but MaxProb vs.~F1 score was negative ($\rho \approx -0.33$), with FPR showing the largest signal (MaxProb vs.~FPR $\rho \approx -0.66$).
At the file level, correlations between UQ and performance were the largest in the benchmark (MaxProb vs.~AUC $\rho \approx 0.79$, MaxProb vs.~F1 score $\rho \approx 0.70$, MaxProb vs.~FPR $\rho \approx -0.70$).
\end{findingbox}

\subsection{RQ2: \rqtwo}
\label{sec:rq2}

RQ2 examined whether predictive performance served as evidence of calibration.
We analyzed this question in two steps.
First, RQ2.1 correlated six performance metrics with three calibration metrics across classifier and dataset observations to test whether high performance was associated with lower calibration error.
Second, RQ2.2 compared calibration profiles across classifier categories to examine whether calibration errors differed by probability-generation mechanism.
Together, these analyses treated predictive performance and calibration as separate evaluation dimensions: performance metrics describe how well classifiers rank or label modules, whereas calibration metrics describe whether predicted probabilities match empirical defect frequencies.

\subsubsection{RQ2.1: Is High Performance Correlated with Low Calibration Error?}
\label{sec:rq2_1}

\begin{figure}[!tbp]
\centering
\includegraphics[width=\linewidth]{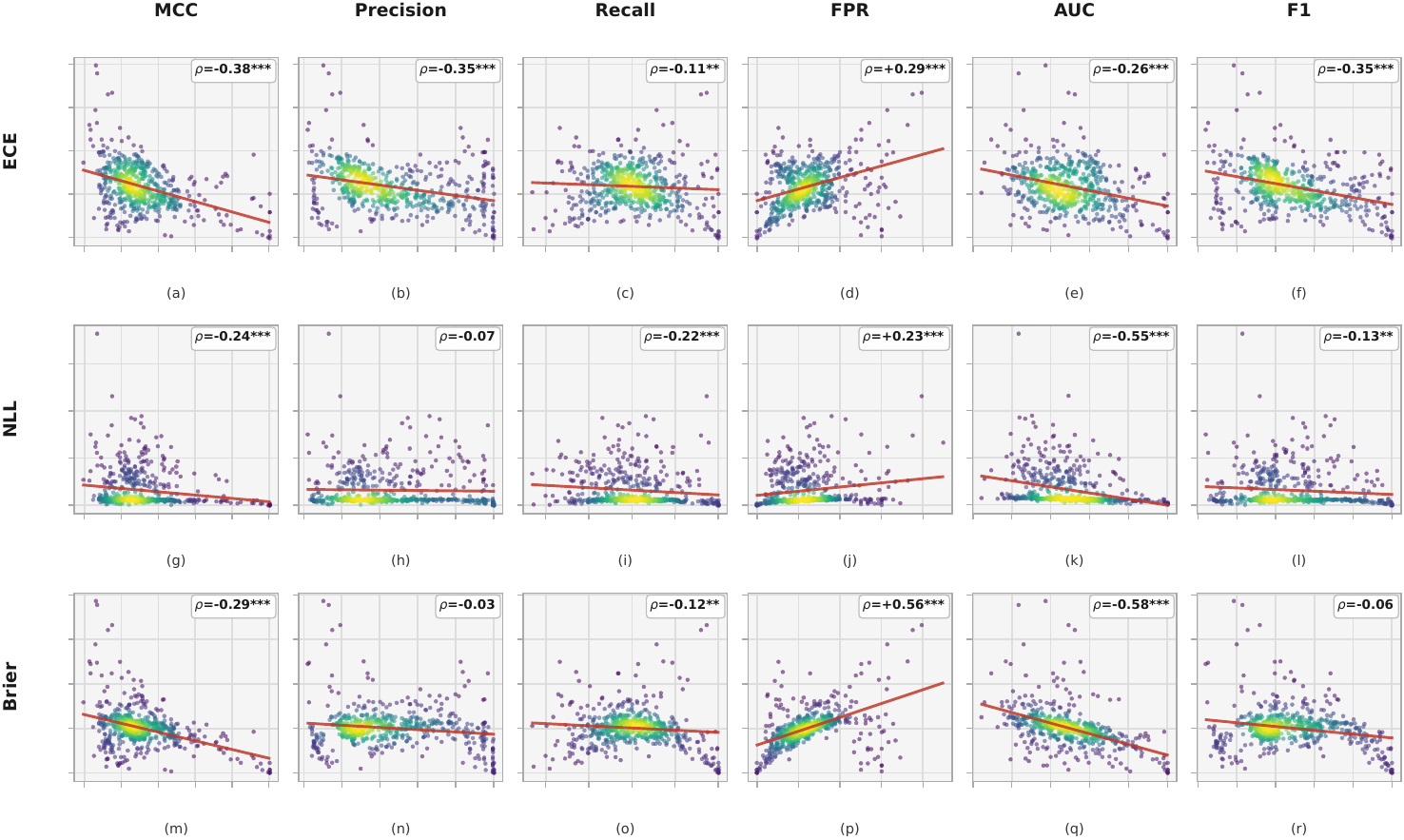}
\caption{\textbf{The correlation between performance and calibration across all metric pairs.} Rows correspond to calibration metrics (ECE, NLL, Brier); columns correspond to performance metrics (MCC, Precision, Recall, FPR, AUC, F1 score). Each panel shows all classifier and dataset observations; color indicates local density; the red line is a linear fit; annotation reports Spearman $\rho$. The largest correlations are AUC vs.~Brier ($\rho = -0.58$), FPR vs.~Brier ($\rho = +0.56$), AUC vs.~NLL ($\rho = -0.55$), and MCC vs.~ECE ($\rho = -0.38$).}
\Description{A 3x6 grid of scatter plots showing all pairwise correlations between six performance metrics and three calibration metrics. Most panels show negative trends with considerable vertical spread.}
\label{fig:performance_calibration_matrix}
\end{figure}

We computed Spearman correlations between each performance metric (MCC, Precision, Recall, FPR, AUC, F1 score) and each calibration metric (ECE, NLL, Brier Score) across all classifier and dataset pairs.
Figure~\ref{fig:performance_calibration_matrix} summarizes all 18 pairwise correlations.
Each panel represents one comparison between performance and calibration across all classifier and dataset observations.
The horizontal axis is a performance metric, and the vertical axis is a calibration error metric.
Darker regions indicate where more observations concentrate, the red line shows the fitted trend, and the annotation gives Spearman $\rho$.
Because ECE, NLL, and Brier Score are error metrics, lower values indicate better calibration.
Thus, a negative correlation for MCC, AUC, Precision, Recall, or F1 score means that better performance tends to coincide with lower calibration error.
For FPR, the interpretation reverses because lower FPR is better; a positive correlation means that higher false positive rates tend to coincide with larger calibration error.

\begin{compactitem}
    \item \textbf{MCC was a moderate indicator of calibration error.}
MCC correlated with ECE at $\rho = -0.38$.

    \item \textbf{AUC was more strongly correlated with Brier Score than with ECE.}
AUC vs.~Brier Score reached $\rho = -0.58$, one of the largest correlations in the matrix.
This was expected because the Brier Score jointly measures calibration and discrimination, whereas ECE measures calibration alone.
AUC vs.~ECE was weaker ($\rho = -0.26$), indicating that higher discrimination does not necessarily translate into calibrated probabilities.

    \item \textbf{FPR was positively correlated with calibration error.}
FPR vs.~ECE was $\rho = +0.29$, and FPR vs.~Brier Score was $\rho = +0.56$.
Classifiers that produced many false positives tended to assign high positive-class probabilities to negative modules.
This behavior increased the Brier Score and could increase ECE.
This pattern is consistent with Finding~1, where UQ has its largest performance correlation with FPR.
\end{compactitem}

\begin{findingbox}
\textbf{Finding 4 (WPDP):} Predictive performance and calibration were correlated but distinct.
MCC correlated negatively with ECE ($\rho = -0.38$), and AUC correlated negatively with Brier Score ($\rho = -0.58$), so better-performing classifiers tended to have lower calibration error on average.
However, the correlation was moderate, and many runs with high performance had large calibration error.
FPR was positively correlated with calibration error ($\rho = +0.56$ vs.~Brier), indicating that classifiers with high false positive rates also tended to have larger probability errors under the Brier Score.
In practice, calibration needs to be checked before deployment.
\end{findingbox}

\subsubsection{RQ2.2: Calibration by Classifier Category}
\label{sec:rq2_2}

Finding~4 showed that performance and calibration had moderate correlations.
We next examined whether calibration itself varied systematically across classifier categories.

\begin{figure}[!tbp]
\centering
\includegraphics[width=\linewidth]{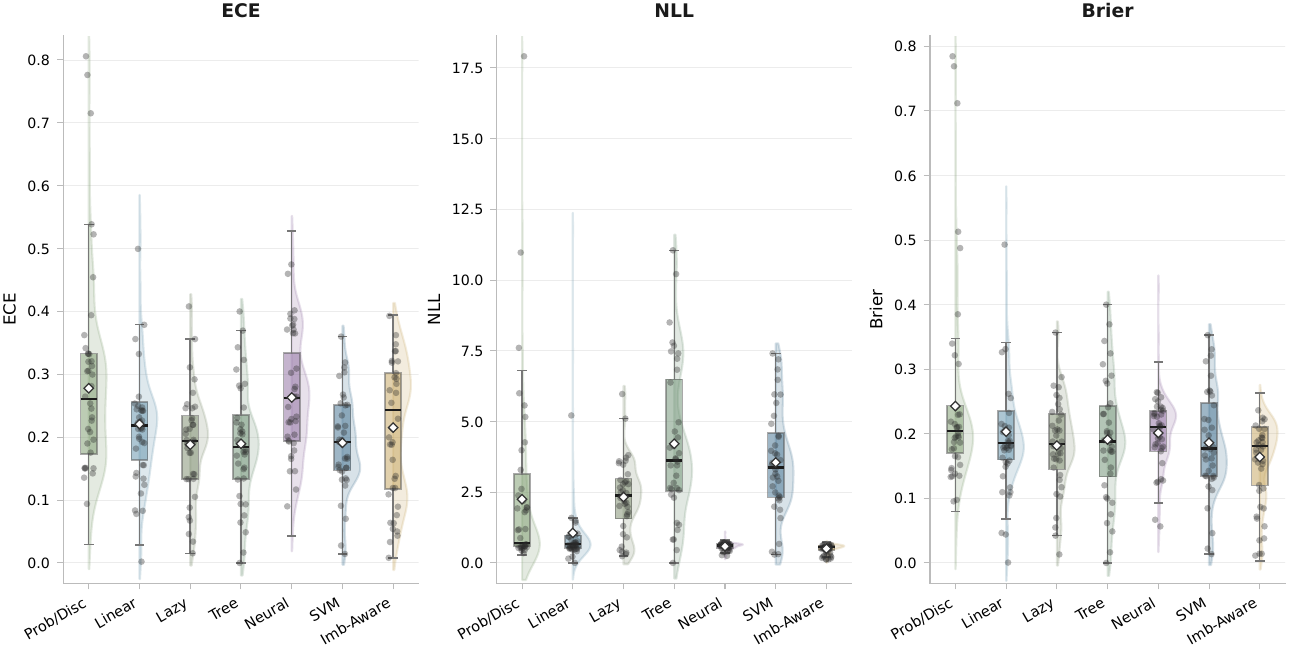}
\caption{\textbf{Calibration distribution by classifier category.} Panels show ECE, NLL, and Brier Score. Each panel displays the distribution of dataset-level calibration errors for all valid classifier and dataset runs within a category: half violin density, box plot, individual points, and mean marker. The categories follow the order of mean ECE. Low ECE does not always coincide with low NLL: KNN and CART have low ECE but high NLL due to occasional extreme probability errors.}
\Description{Three panels of half violin box plots showing calibration error distributions for the seven classifier categories listed in Table~1. The Ensemble category has the lowest aggregate NLL/Brier values. Lazy Learning and Decision Tree have low ECE but high NLL. Probabilistic Model and Neural Network have the highest ECE.}
\label{fig:category_calibration_summary}
\end{figure}

For each classifier, we computed mean ECE, NLL, and Brier Score across all datasets.
We grouped classifiers according to the categories listed in Table~\ref{tab:model_categories}.
Figure~\ref{fig:category_calibration_summary} shows the calibration profile of each classifier category across the three metrics.
Each panel represents one calibration error metric across classifier categories.
The horizontal axis lists the classifier categories, and the vertical axis reports the calibration error value.
The violin shape shows the distribution of observations within a category, the box summarizes the median and interquartile range, the points show individual classifier and dataset runs, and the mean marker shows the category average.
Because ECE, NLL, and Brier Score are all error metrics, lower positions indicate better calibration in each panel.
Differences across panels are important: a category can have low ECE but high NLL, which means average bin error is small while occasional extreme probability errors remain large.

\begin{compactitem}
    \item \textbf{Probabilistic Model: the largest aggregate calibration error.}
This category had the largest aggregate calibration error (ECE $= 0.278$, NLL $= 2.166$, Brier Score $= 0.242$).
Naive Bayes and QDA were the main contributors: Naive Bayes had ECE $= 0.302$, NLL $= 4.081$, and Brier Score $= 0.295$; QDA had ECE $= 0.299$, NLL $= 3.473$, and Brier Score $= 0.291$.
Gaussian Process Classifier had lower calibration error (ECE $= 0.239$, NLL $= 0.538$) than the other members of this category, but its calibration remained weaker than that of the Linear Model and Ensemble categories.
The independence assumptions in Naive Bayes and the distributional assumptions in QDA can push probability mass toward extremes.
This pattern was consistent with their larger calibration errors on these SDP data.

    \item \textbf{Linear Model: moderate calibration error with internal heterogeneity.}
This category showed the widest internal spread.
Logistic Regression had lower calibration error (ECE $= 0.192$, NLL $= 0.545$), but SGDClassifier (ECE $= 0.237$, NLL $= 1.815$) and PassiveAggressiveClassifier (ECE $= 0.242$, NLL $= 0.763$) increased the category average.
These two classifiers also had high Recall in the category ($0.625$ and $0.610$).
This suggests that recall-oriented linear methods can produce positive-class probability errors.

    \item \textbf{Lazy Learning: low ECE but high NLL.}
KNN had mean ECE $= 0.183$ and Brier Score $= 0.182$, but its NLL was $2.212$.
Its distance-based probability estimates made bin-level averages appear low while occasional extreme probability errors remained large under log loss.

    \item \textbf{Decision Tree: low ECE but high NLL.}
CART had mean ECE $= 0.195$ and Brier Score $= 0.196$, but NLL reached $4.423$, the highest among all 16 classifiers.
Its leaf node frequency probabilities were nearly binary, causing extreme log loss penalties on misclassified instances.

    \item \textbf{Neural Network: moderate calibration error with internal differentiation.}
This category (MLP 1, MLP 2, MLP Bagging) had mean ECE $= 0.255$, NLL $= 0.576$, and Brier Score $= 0.197$.
MLP 1 and MLP 2 produced relatively high bin-level calibration error even though their NLL remained moderate.
MLP Bagging had higher ECE ($0.246$) but low NLL ($0.546$) and Brier Score ($0.184$).
It therefore had a different calibration profile from the single MLP classifiers.
The three neural classifiers thus displayed heterogeneous calibration behavior despite sharing the same softmax probability mechanism.

    \item \textbf{SVM-Based: extreme errors under log loss.}
SVM Bagging had low ECE ($0.199$) but high NLL ($3.717$), indicating occasional extreme probability errors.
Its probability estimates could produce extreme values near 0 or 1 that inflated NLL but averaged out in bin-level ECE.

    \item \textbf{Ensemble: the lowest aggregate NLL and Brier Score.}
This category (BalancedRandomForest, EasyEnsemble, RUSBoost) had mean ECE $= 0.214$, NLL $= 0.509$, and Brier Score $= 0.164$.
BalancedRandomForest had the lowest ECE ($0.100$) and also had the lowest Brier Score ($0.113$).
EasyEnsemble and RUSBoost had higher ECE ($0.266$ and $0.277$), but their NLL and Brier values remained moderate.
\end{compactitem}

\begin{findingbox}
\textbf{Finding 5 (WPDP):} Calibration error varied by classifier category.
For instance, the Ensemble category had the lowest aggregate NLL/Brier profile, with BalancedRandomForest showing the lowest individual ECE ($0.100$).
SVM-Based classifiers had low ECE but high NLL because extreme probability estimates inflated log-loss penalties.
\end{findingbox}

\subsection{RQ3: \rqthree}
\label{sec:rq3_cpdp}

RQ1 and RQ2 examined WPDP, where training and test data came from the same project.
CPDP supports prediction when a new project lacks historical defect data.
RQ3 asked how UQ related to performance and calibration under CPDP, and whether those correlations changed from WPDP to CPDP.

We evaluated CPDP using a within-collection Direct Transfer protocol~\cite{zimmermann2009cross,nam2013transfer,nam2015heterogeneous,amasaki2018cross}.
Within each included collection, one dataset served as the source project, and a different dataset from the same collection served as the target project.
We aligned source and target features by common feature names and retained only source-target pairs with sufficient feature overlap.
We excluded ReLink because its four datasets formed two feature-incompatible subgroups.
The two subgroups had no overlapping features, and each subgroup contained only two datasets, leaving too few source-target pairs for stable CPDP analysis.
The CPDP experiments therefore focused on AEEEM (5 datasets), NASA (12 datasets), and PROMISE (15 datasets).

\subsubsection{RQ3.1: How Is UQ Related to Performance and Calibration Under CPDP?}
\label{sec:rq3_1}

\noindent\textbf{Experimental Design.}
We generated CPDP results using Direct Transfer for all ordered source $\rightarrow$ target permutations within AEEEM, NASA, and PROMISE.
The evaluation used the 16 classifiers for all ordered source-target pairs.
For each dataset collection, we computed the Spearman correlation between each UQ metric and each CPDP target metric.
The target metrics included six performance metrics (MCC, Precision, Recall, FPR, AUC, and F1 score) and three calibration metrics (ECE, NLL, and Brier Score).

\noindent\textbf{Result.}
The CPDP correlations showed two patterns.
This subsection reads the CPDP correlations directly; RQ3.2 then compares them with WPDP.

\begin{compactitem}
    \item \textbf{UQ and performance under CPDP.}
Performance metric correlations under CPDP remained specific to the metric and dataset collection rather than global.
In NASA, the largest visible performance correlations involved AUC: MaxProb vs.~AUC was negative ($\rho=-0.342$), while entropy-based AUC correlations were positive.
AEEEM and PROMISE did not show a uniform CPDP performance pattern across all UQ metrics and performance metrics.
Thus, CPDP did not turn UQ into a general performance signal.

    \item \textbf{UQ and calibration under CPDP.}
Calibration metric correlations were more sensitive to the dataset collection.
In AEEEM, MaxProb vs.~Brier reached $\rho=0.328$ under CPDP.
In PROMISE, MaxProb vs.~ECE reached $\rho=0.234$, and the mean absolute correlation for calibration pairs was $0.482$ under CPDP.
NASA showed weaker CPDP calibration correlations: its two Brier Score cases were close to zero.
These results indicated that CPDP can change how confidence-based UQ relates to calibration error.
\end{compactitem}

\begin{findingbox}
\textbf{Finding 6 (CPDP):} Under CPDP, UQ was related to both performance and calibration, but the correlation was conditional on the metric and dataset collection.
Performance relations concentrated in selected metric pairs rather than all performance metrics.
Calibration relations showed stronger collection-specific behavior, especially in AEEEM and PROMISE.
For SDP practice, CPDP evaluation should report UQ correlations with both performance metrics and calibration metrics.
\end{findingbox}

\subsubsection{RQ3.2: How Do UQ Relationships Change from WPDP to CPDP?}
\label{sec:rq3_2}

\noindent\textbf{Experimental Design.}
Using the same (UQ, target metric) correlations, we compared the CPDP Spearman correlation with the corresponding WPDP Spearman correlation within each dataset collection.
We used \emph{correlation direction reversal} as a signal of transfer stability.
We counted a reversal when the WPDP and CPDP correlations pointed in opposite directions.

\noindent\textbf{Result.}
Table~\ref{tab:rq3_corr_summary} summarizes the comparison between WPDP and CPDP numerically.
Reading the table along the WPDP-to-CPDP axis separates what transferred from what did not: UQ--performance correlations were largely stable in direction, with no performance reversals in NASA or PROMISE, whereas UQ--calibration correlations reversed direction in a collection-dependent way.
Three collection-level changes made this split concrete.

\begin{table}[!htbp]
\centering
\footnotesize
\setlength{\tabcolsep}{3pt}
\caption{\textbf{Summary of UQ correlation changes from WPDP to CPDP.} Each row summarizes 45 pairs formed by five UQ metrics and nine target metrics. The reversal column reports total correlation direction reversals, with performance/calibration counts in parentheses.}
\label{tab:rq3_corr_summary}
\begin{tabularx}{\linewidth}{@{}lcccX@{}}
\toprule
\textbf{Collection} & \textbf{WPDP $|\rho|$} & \textbf{CPDP $|\rho|$} & \textbf{Rev. (Perf/Cal)} & \textbf{Main pattern} \\
\midrule
AEEEM & 0.251 & 0.222 & 15 (5/10) & Calibration changes dominate, with additional performance reversals. \\
NASA & 0.227 & 0.279 & 2 (0/2) & Mostly stable; only weak Brier Score reversals appear. \\
PROMISE & 0.141 & 0.234 & 5 (0/5) & Calibration direction reversals dominate, especially for ECE. \\
\bottomrule
\end{tabularx}
\end{table}

\begin{compactitem}
    \item \textbf{AEEEM: calibration correlations were unstable under CPDP.}
AEEEM had 15 descriptive direction reversals out of 45 UQ and target metric pairs.
Ten involved calibration metrics; for example, MaxProb vs.~Brier changed from $\rho = -0.177$ in WPDP to $\rho = 0.328$ under CPDP.
The mean absolute correlation decreased from $0.251$ in WPDP to $0.222$ under CPDP.
This indicated that, in AEEEM, the correlation between probability concentration and calibration error changed under transfer.

    \item \textbf{NASA: correlations increased with only weak Brier Score reversals.}
NASA had 2 direction reversals, both weak calibration pairs (Predictive Entropy/DeepGini vs.~Brier Score) where the WPDP correlation was slightly above $0.1$ and the CPDP correlation was close to zero.
The mean absolute correlation increased from $0.227$ in WPDP to $0.279$ under CPDP.
The largest changes involved performance metrics: the MaxProb vs.~AUC correlation became more negative ($-0.070$ to $-0.342$), and the entropy-based AUC correlations became more positive.
Thus, CPDP did not remove the NASA correlations between UQ and target metrics.
Several correlations increased, while only the two weak Brier Score pairs crossed the reversal criterion.

    \item \textbf{PROMISE: calibration direction reversals concentrated in calibration metrics.}
PROMISE had 5 direction reversals, all involving calibration metrics.
For example, MaxProb vs.~ECE changed from $\rho = -0.208$ in WPDP to $\rho = 0.234$ under CPDP.
The mean absolute correlation increased from $0.141$ in WPDP to $0.234$ under CPDP, mainly because calibration pairs increased from $0.271$ to $0.482$.
This indicated that CPDP can reverse the correlation between confidence and calibration error even when correlations between UQ and performance remain weak.
\end{compactitem}

\begin{findingbox}
\textbf{Finding 7 (CPDP):} UQ correlations changed from WPDP to CPDP in a collection-dependent manner.
AEEEM and PROMISE showed calibration metric direction reversals, while NASA showed only two weak Brier Score reversals.
Practitioners should not transfer the direction and magnitude of UQ correlations from WPDP to CPDP without target collection checks.
\end{findingbox}

\section{Discussion and Implications}
\label{sec:discussion}

\subsection{Summary of Findings}
\label{sec:summary}

The seven findings showed that probability-based UQ is informative in SDP, but only under explicit evaluation conditions.
The meaning of a UQ score depends on the target metric, classifier category, dataset collection, prediction granularity, and transfer setting.
Table~\ref{tab:summary} summarizes this interpretation across the three empirical dimensions examined in this study.
The first two dimensions are established under WPDP, whereas the third examines how these relationships transfer to CPDP.

\begin{table}[!htbp]
\centering
\small
\caption{Summary of findings by analytical dimension and evidence level.
}
\vspace{-10pt}
\label{tab:summary}
\begin{tabularx}{\textwidth}{@{}>{\raggedright\arraybackslash}p{2.7cm} >{\raggedright\arraybackslash}p{2.1cm} >{\raggedright\arraybackslash}X >{\raggedright\arraybackslash}p{2.8cm}@{}}
\toprule
\textbf{Dimension} & \textbf{Findings} & \textbf{Key Evidence} & \textbf{Level} \\
\midrule
UQ and performance
& Findings~1, 2, 3 (WPDP)
& UQ was strongly correlated with FPR and AUC. Correlations with MCC, Precision, Recall, and F1 score were weak or unstable. Alignment also varied across classifier categories and dataset groups
& Global, category, dataset collection \\
\midrule
Performance and calibration
& Findings~4, 5 (WPDP)
& Predictive performance had a moderate correlation with calibration error, but high discrimination could coexist with large ECE, NLL, or Brier Score. Calibration profiles differed by classifier category
& Metric, category \\
\midrule
Cross-project transfer
& Findings~6, 7 (CPDP)
& UQ remained correlated with selected performance and calibration metrics under CPDP. Several correlations changed magnitude or direction, especially calibration correlations in AEEEM and PROMISE
& Dataset collection \\
\bottomrule
\end{tabularx}
\end{table}

This summary clarifies three points about UQ in SDP, each corresponding to a knowledge gap identified in Section~\ref{sec:study_motivation}.
First, regarding Gap~1, UQ has metric-specific implications rather than serving as a general measure of classifier quality.
The global WPDP analysis showed consistent correlations mainly for FPR and AUC.
The category and dataset analyses further showed that the same UQ metric can become weak, inverted, or collection-specific under different settings.
Thus, a high or low uncertainty score is not sufficient on its own.
It needs to be interpreted together with the performance metric and the dataset collection.

Second, regarding Gap~2, calibration remains a separate evaluation dimension that cannot be inferred from predictive performance alone.
Finding~4 showed that better predictive performance was correlated with lower calibration error on average.
However, the correlation was moderate, and many high-performing runs still showed large calibration error.
Finding~5 further showed that calibration profiles varied by classifier category.
For example, CART and SVM Bagging had low ECE but high NLL because extreme probability estimates created large log loss penalties.
Predicted probabilities in SDP therefore require calibration evaluation when they are analyzed as probabilities rather than only as class labels.

Third, regarding Gap~3, cross-project transfer changes how UQ should be read.
RQ3 showed that UQ remained related to selected performance and calibration metrics under CPDP, but these relations were collection-dependent.
AEEEM and PROMISE showed calibration direction reversals, while NASA showed only two weak Brier Score reversals.
WPDP evidence is therefore useful as an initial reference, but it does not determine how UQ behaves after transfer to a target collection.

These results complement prior UQ studies based on Bayesian neural networks and deep ensembles~\cite{gal2016dropout,lakshminarayanan2017simple} by showing that probability outputs from classical SDP classifiers also provide measurable but context-dependent implications for uncertainty interpretation.
They also relate recent SDP calibration studies~\cite{shahini2024empirical,shahini2025calibration} to foundational CPDP research~\cite{zimmermann2009cross,nam2013transfer} and broader calibration work~\cite{dal2015calibrating,van2022harm}.
Together, these findings indicate that calibration, predictive performance, and transfer behavior should be reported as separate evaluation dimensions.

\subsection{Implications for Evaluation and Reporting}
\label{sec:implications}

The findings suggest the following practices for SDP studies that use probability-based UQ.

\textbf{1. Do not use UQ as a proxy for MCC or F1 score.}
UQ showed consistent correlations with FPR and AUC, but its correlations with MCC, F1 score, Precision, and Recall were weak or unstable across classifiers and dataset collections (Findings~1 and~3).
In practice, MCC and F1 score should be measured directly rather than inferred from uncertainty scores, because UQ does not reliably reflect these metrics.

\textbf{2. Select UQ-informative classifiers deliberately.}
UQ signals were informative for AUC and FPR across most classifiers, but were weak for threshold-dependent metrics and absent for a few classifiers even on AUC (Finding~2).
Neural networks (MLP~1, MLP-Bagging) and imbalance-aware ensembles (EasyEnsemble, RUSBoost) produced strong and consistent UQ--AUC alignment ($\ge 0.70$, with narrow min--max ranges), whereas CART ($-0.02$) and SGDClassifier ($0.00$) produced near-zero alignment.
Practitioners who intend to use UQ for model selection or performance monitoring should rely on it only for ranking-oriented metrics (AUC and FPR) and should avoid classifiers with degenerate probability outputs (e.g., CART, SGDClassifier), whose UQ carries no usable signal.

\textbf{3. Report multiple calibration metrics, not ECE alone.}
CART and SVM Bagging had low ECE but high NLL because their extreme probability estimates incurred large log-loss penalties that bin-averaged ECE masked (Finding~5).
Future SDP studies that analyze predicted probabilities should report at least two calibration metrics (e.g., ECE together with NLL or Brier Score) to avoid misleading conclusions about probability quality.

\textbf{4. Treat calibration as independent of discrimination.}
High AUC or MCC does not guarantee well-calibrated probabilities (Finding~4).
Practitioners who use predicted probabilities for inspection prioritization, threshold tuning, or cost-sensitive decision-making should evaluate calibration independently, because a classifier that ranks modules correctly may still assign inaccurate defect probabilities.

\textbf{5. Account for dataset collection effects when interpreting UQ.}
Even within the same prediction granularity, UQ profiles differed across collections: AEEEM showed near-zero UQ--performance signals while PROMISE showed weak but detectable correlations (Finding~3).
Future UQ studies should report results separately for each dataset collection, because aggregate correlations can mask opposing patterns across collections (e.g., near-zero signals in AEEEM vs.\ weak positive signals in PROMISE).

\textbf{6. Re-evaluate UQ--calibration correlations after cross-project transfer.}
Under CPDP, several calibration correlations changed direction in AEEEM and PROMISE, while NASA largely preserved its WPDP patterns (Findings~6 and~7).
CPDP practitioners should not assume that UQ--calibration correlations observed in within-project settings transfer unchanged; re-evaluation on the target collection is necessary before using UQ for deployment decisions.

\section{Threats to Validity}
\label{sec:threats}

\noindent\textbf{Internal Validity.}
\label{sec:internal_validity}
The main internal threat is leakage from preprocessing and repeated evaluation.
The experimental protocol separates training and evaluation before scaling.
In WPDP, we performed the training and test split before scaling; we fitted the scaler on the training subset and then applied it to the test subset.
In CPDP, we fitted the scaler on the source project and then applied it to the target project.
This design keeps target project information out of model fitting and preprocessing.
To reduce random variation from training and test splits and stochastic classifiers, all WPDP and CPDP experiments used the repeated run protocol described in Section~\ref{sec:experimental_protocol}.
The correlations therefore summarized repeated observations rather than a single split or seed.

\noindent\textbf{External Validity.}
\label{sec:external_validity}
The benchmark includes 36 datasets from four collections (AEEEM, NASA, PROMISE, and ReLink), spanning multiple programming languages and application domains.
However, all datasets come from public repositories commonly used in SDP research.
The results may not generalize to proprietary industrial systems with different feature distributions, labeling practices, or defect reporting processes.
The classifier set covers 16 classical classifiers that are common in SDP studies, but it does not include newer code representation models or pretrained language models for code.

Another external threat is the coupling between prediction granularity and dataset collection.
In this benchmark, class-level datasets come from AEEEM and PROMISE, function/method-level datasets from NASA, and file-level datasets from ReLink.
The granularity-aware analysis may therefore reflect collection-specific data collection, feature definitions, or label quality rather than prediction granularity alone.
We addressed this threat by interpreting Finding~3 as collection and granularity heterogeneity rather than as a causal effect of granularity.
The CPDP analysis is also limited to AEEEM, NASA, and PROMISE because ReLink lacks sufficient feature-compatible source-target pairs under the within-collection protocol.
CPDP conclusions therefore apply to feature-compatible collection-level transfers and should not be generalized to file-level ReLink transfer.

\noindent\textbf{Construct Validity.}
\label{sec:construct_validity}
The UQ measures in this study are five standard scores derived from predictive probabilities.
In binary classification, these scores are closely related functions of the predicted class probabilities, so we do not interpret them as five fully independent uncertainty constructs.
For classifiers without native probabilistic outputs, probability estimates depend on post hoc transformations or model-specific scoring functions, which can affect calibration behavior.
The calibration metrics (ECE, NLL, and Brier Score) also summarize different aspects of probability error.
ECE reports average bin-level error, NLL emphasizes extreme probability errors, and Brier Score measures squared probability error.
As a result, the study characterizes dataset-level relationships among UQ, performance, and calibration rather than instance-level uncertainty explanations.
Different binning schemes for ECE may also change absolute ECE values, although the paper focuses mainly on relationships and category-level patterns.

\section{Conclusion}
\label{sec:conclusion}

This study set out to clarify what probability-based uncertainty quantification means in software defect prediction.
Across 16 representative classifiers and 36 benchmark datasets, we examined relationships among UQ, predictive performance, and probability calibration.
We also tested whether these relationships changed under CPDP using 32 feature-compatible datasets from AEEEM, NASA, and PROMISE.

The findings indicate that UQ is informative only when its evaluation context is explicit.
In WPDP, UQ--performance correlations were weak at the global level except for selected metric pairs, especially FPR and AUC.
These correlations also varied across classifier categories and granularity-aware dataset groups.
Most classifiers showed moderate-to-strong AUC alignment, led by neural networks and imbalance-aware ensembles, whereas only CART and SGDClassifier showed near-zero alignment.
Predictive performance and calibration were related but not interchangeable.
High discrimination could coexist with large calibration error, and classifier categories differed across ECE, NLL, and Brier Score.
Under CPDP, UQ remained related to selected performance and calibration metrics.
However, changes from WPDP to CPDP were collection-dependent.
AEEEM and PROMISE showed calibration direction reversals, whereas NASA showed only two weak Brier Score reversals.

These results refine how probability scores should be reported in SDP studies.
UQ should be linked to the specific performance metric it is intended to inform, and MCC or F1 score should not be inferred from uncertainty scores.
Classifier choice matters: most classifiers provide informative UQ signals for AUC and FPR, but a few with degenerate probabilities (CART, SGDClassifier) do not.
Calibration should be evaluated with multiple metrics and independently of discrimination.
Results should be reported at the collection level, and UQ--calibration correlations should be re-evaluated after cross-project transfer.
These conclusions are bounded by the classical probability-output classifiers and public tabular datasets studied here.
Future work can examine whether the same relationship patterns hold for Bayesian neural networks, deep ensembles, pretrained code representations, and other modern feature spaces.

\bibliographystyle{ACM-Reference-Format} \bibliography{sample-base}

@inproceedings{omri2020deep,
  title={Deep learning for software defect prediction: A survey},
  author={Omri, Safa and Sinz, Carsten},
  booktitle={Proceedings of the IEEE/ACM 42nd international conference on software engineering workshops},
  pages={209--214},
  year={2020}
}

@article{goyal2025current,
  title={Current Trends in Class Imbalance Learning for Software Defect Prediction},
  author={Goyal, Somya R},
  journal={IEEE Access},
  year={2025},
  publisher={IEEE}
}

@article{menzies2010defect,
  title={Defect prediction from static code features: current results, limitations, new approaches},
  author={Menzies, Tim and Milton, Zach and Turhan, Burak and Cukic, Bojan and Jiang, Yue and Bener, Ayse},
  journal={Automated Software Engineering},
  volume={17},
  number={4},
  pages={375--407},
  year={2010},
  publisher={Springer}
}

@article{elhassan2016classification,
  title={Classification of imbalance data using tomek link (t-link) combined with random under-sampling (rus) as a data reduction method},
  author={Elhassan, T and Aljurf, M and others},
  journal={Global J Technol Optim S},
  volume={1},
  number={S1},
  year={2016}
}

@inproceedings{he2008adasyn,
  title={ADASYN: Adaptive synthetic sampling approach for imbalanced learning},
  author={He, Haibo and Bai, Yang and Garcia, Edwardo A and Li, Shutao},
  booktitle={2008 IEEE international joint conference on neural networks (IEEE world congress on computational intelligence)},
  pages={1322--1328},
  year={2008},
  organization={Ieee}
}

@inproceedings{han2005borderline,
  title={Borderline-SMOTE: a new over-sampling method in imbalanced data sets learning},
  author={Han, Hui and Wang, Wen-Yuan and Mao, Bing-Huan},
  booktitle={International conference on intelligent computing},
  pages={878--887},
  year={2005},
  organization={Springer}
}

@article{matloob2021software,
  title={Software defect prediction using ensemble learning: A systematic literature review},
  author={Matloob, Faseeha and Ghazal, Taher M and Taleb, Nasser and Aftab, Shabib and Ahmad, Munir and Khan, Muhammad Adnan and Abbas, Sagheer and Soomro, Tariq Rahim},
  journal={IEEe Access},
  volume={9},
  pages={98754--98771},
  year={2021},
  publisher={IEEE}
}

@article{zhao2023systematic,
  title={A systematic survey of just-in-time software defect prediction},
  author={Zhao, Yunhua and Damevski, Kostadin and Chen, Hui},
  journal={ACM Computing Surveys},
  volume={55},
  number={10},
  pages={1--35},
  year={2023},
  publisher={ACM New York, NY}
}

@article{abdar2021review,
  title={A review of uncertainty quantification in deep learning: Techniques, applications and challenges},
  author={Abdar, Moloud and Pourpanah, Farhad and Hussain, Sadiq and Rezazadegan, Dana and Liu, Li and Ghavamzadeh, Mohammad and Fieguth, Paul and Cao, Xiaochun and Khosravi, Abbas and Acharya, U Rajendra and others},
  journal={Information fusion},
  volume={76},
  pages={243--297},
  year={2021},
  publisher={Elsevier}
}

@inproceedings{hassan2009predicting,
  title={Predicting faults using the complexity of code changes},
  author={Hassan, Ahmed E},
  booktitle={2009 IEEE 31st international conference on software engineering},
  pages={78--88},
  year={2009},
  organization={IEEE}
}

@article{li2018progress,
  title={Progress on approaches to software defect prediction},
  author={Li, Zhiqiang and Jing, Xiao-Yuan and Zhu, Xiaoke},
  journal={Iet Software},
  volume={12},
  number={3},
  pages={161--175},
  year={2018},
  publisher={Wiley Online Library}
}

@inproceedings{agrawal2018better,
  title={Is" better data" better than" better data miners"? on the benefits of tuning SMOTE for defect prediction},
  author={Agrawal, Amritanshu and Menzies, Tim},
  booktitle={Proceedings of the 40th International Conference on Software engineering},
  pages={1050--1061},
  year={2018}
}

@inproceedings{nixon2019measuring,
  title={Measuring calibration in deep learning.},
  author={Nixon, Jeremy and Dusenberry, Michael W and Zhang, Linchuan and Jerfel, Ghassen and Tran, Dustin},
  booktitle={CVPR workshops},
  volume={2},
  number={7},
  year={2019}
}

@article{minderer2021revisiting,
  title={Revisiting the calibration of modern neural networks},
  author={Minderer, Matthias and Djolonga, Josip and Romijnders, Rob and Hubis, Frances and Zhai, Xiaohua and Houlsby, Neil and Tran, Dustin and Lucic, Mario},
  journal={Advances in neural information processing systems},
  volume={34},
  pages={15682--15694},
  year={2021}
}

@article{wu2025uncertaintyzoo,
  title={UncertaintyZoo: A Unified Toolkit for Quantifying Predictive Uncertainty in Deep Learning Systems},
  author={Wu, Xianzong and Li, Xiaohong and Quan, Lili and Hu, Qiang},
  journal={arXiv preprint arXiv:2512.06406},
  year={2025}
}

@article{blagus2013smote,
  title={SMOTE for high-dimensional class-imbalanced data},
  author={Blagus, Rok and Lusa, Lara},
  journal={BMC bioinformatics},
  volume={14},
  number={1},
  pages={106},
  year={2013},
  publisher={Springer}
}

@article{canfora2015defect,
  title={Defect prediction as a multiobjective optimization problem},
  author={Canfora, Gerardo and Lucia, Andrea De and Penta, Massimiliano Di and Oliveto, Rocco and Panichella, Annibale and Panichella, Sebastiano},
  journal={Software Testing, Verification and Reliability},
  volume={25},
  number={4},
  pages={426--459},
  year={2015},
  publisher={Wiley Online Library}
}

@inproceedings{wu2011relink,
  title={Relink: recovering links between bugs and changes},
  author={Wu, Rongxin and Zhang, Hongyu and Kim, Sunghun and Cheung, Shing-Chi},
  booktitle={Proceedings of the 19th ACM SIGSOFT symposium and the 13th European conference on Foundations of software engineering},
  pages={15--25},
  year={2011}
}

@article{catal2009systematic,
  title={A systematic review of software fault prediction studies},
  author={Catal, Cagatay and Diri, Banu},
  journal={Expert systems with applications},
  volume={36},
  number={4},
  pages={7346--7354},
  year={2009},
  publisher={Elsevier}
}

@article{hall2011systematic,
  title={A systematic literature review on fault prediction performance in software engineering},
  author={Hall, Tracy and Beecham, Sarah and Bowes, David and Gray, David and Counsell, Steve},
  journal={IEEE Transactions on Software Engineering},
  volume={38},
  number={6},
  pages={1276--1304},
  year={2011},
  publisher={IEEE}
}

@article{de2016comparing,
  title={Comparing the Pearson and Spearman correlation coefficients across distributions and sample sizes: A tutorial using simulations and empirical data.},
  author={De Winter, Joost CF and Gosling, Samuel D and Potter, Jeff},
  journal={Psychological methods},
  volume={21},
  number={3},
  pages={273},
  year={2016},
  publisher={American Psychological Association}
}

@article{wan2024data,
  title={Data complexity: a new perspective for analyzing the difficulty of defect prediction tasks},
  author={Wan, Xiaohui and Zheng, Zheng and Qin, Fangyun and Lu, Xuhui},
  journal={ACM Transactions on Software Engineering and Methodology},
  volume={33},
  number={6},
  pages={1--45},
  year={2024},
  publisher={ACM New York, NY}
}

@inproceedings{feng2020deepgini,
  title={Deepgini: prioritizing massive tests to enhance the robustness of deep neural networks},
  author={Feng, Yang and Shi, Qingkai and Gao, Xinyu and Wan, Jun and Fang, Chunrong and Chen, Zhenyu},
  booktitle={Proceedings of the 29th ACM SIGSOFT international symposium on software testing and analysis},
  pages={177--188},
  year={2020}
}

@article{breiman1996bagging,
  title={Bagging predictors},
  author={Breiman, Leo},
  journal={Machine learning},
  volume={24},
  number={2},
  pages={123--140},
  year={1996},
  publisher={Springer}
}

@inproceedings{amasaki2018cross,
  title={Cross-version defect prediction using cross-project defect prediction approaches: Does it work?},
  author={Amasaki, Sousuke},
  booktitle={Proceedings of the 14th International Conference on predictive models and data analytics in software engineering},
  pages={32--41},
  year={2018}
}

@article{tantithamthavorn2016empirical,
  title={An empirical comparison of model validation techniques for defect prediction models},
  author={Tantithamthavorn, Chakkrit and McIntosh, Shane and Hassan, Ahmed E and Matsumoto, Kenichi},
  journal={IEEE Transactions on Software Engineering},
  volume={43},
  number={1},
  pages={1--18},
  year={2016},
  publisher={IEEE}
}

@inproceedings{guo2017calibration,
  title={On calibration of modern neural networks},
  author={Guo, Chuan and Pleiss, Geoff and Sun, Yu and Weinberger, Kilian Q},
  booktitle={International conference on machine learning},
  pages={1321--1330},
  year={2017},
  organization={PMLR}
}

@article{fan2019impact,
  title={The impact of mislabeled changes by szz on just-in-time defect prediction},
  author={Fan, Yuanrui and Xia, Xin and Da Costa, Daniel Alencar and Lo, David and Hassan, Ahmed E and Li, Shanping},
  journal={IEEE transactions on software engineering},
  volume={47},
  number={8},
  pages={1559--1586},
  year={2019},
  publisher={IEEE}
}

@article{he2025survey,
  title={A survey on uncertainty quantification methods for deep learning},
  author={He, Wenchong and Jiang, Zhe and Xiao, Tingsong and Xu, Zelin and Li, Yukun},
  journal={ACM Computing Surveys},
  year={2025},
  publisher={ACM New York, NY}
}

@article{lakshminarayanan2017simple,
  title={Simple and scalable predictive uncertainty estimation using deep ensembles},
  author={Lakshminarayanan, Balaji and Pritzel, Alexander and Blundell, Charles},
  journal={Advances in neural information processing systems},
  volume={30},
  year={2017}
}

@inproceedings{niculescu2005predicting,
  title={Predicting good probabilities with supervised learning},
  author={Niculescu-Mizil, Alexandru and Caruana, Rich},
  booktitle={Proceedings of the 22nd international conference on Machine learning},
  pages={625--632},
  year={2005}
}

@inproceedings{li2020understanding,
  title={Understanding the automated parameter optimization on transfer learning for cross-project defect prediction: an empirical study},
  author={Li, Ke and Xiang, Zilin and Chen, Tao and Wang, Shuo and Tan, Kay Chen},
  booktitle={Proceedings of the ACM/IEEE 42nd International Conference on Software Engineering},
  pages={566--577},
  year={2020}
}

@inproceedings{tantithamthavorn2016automated,
  title={Automated parameter optimization of classification techniques for defect prediction models},
  author={Tantithamthavorn, Chakkrit and McIntosh, Shane and Hassan, Ahmed E and Matsumoto, Kenichi},
  booktitle={Proceedings of the 38th international conference on software engineering},
  pages={321--332},
  year={2016}
}

@article{tantithamthavorn2018impact,
  title={The impact of automated parameter optimization on defect prediction models},
  author={Tantithamthavorn, Chakkrit and McIntosh, Shane and Hassan, Ahmed E and Matsumoto, Kenichi},
  journal={IEEE Transactions on Software Engineering},
  volume={45},
  number={7},
  pages={683--711},
  year={2018},
  publisher={IEEE}
}

@inproceedings{d2010extensive,
  title={An extensive comparison of bug prediction approaches},
  author={D'Ambros, Marco and Lanza, Michele and Robbes, Romain},
  booktitle={2010 7th IEEE working conference on mining software repositories (MSR 2010)},
  pages={31--41},
  year={2010},
  organization={IEEE}
}

@article{shepperd2013data,
  title={Data quality: Some comments on the nasa software defect datasets},
  author={Shepperd, Martin and Song, Qinbao and Sun, Zhongbin and Mair, Carolyn},
  journal={IEEE Transactions on software engineering},
  volume={39},
  number={9},
  pages={1208--1215},
  year={2013},
  publisher={IEEE}
}

@inproceedings{jureczko2010towards,
  title={Towards identifying software project clusters with regard to defect prediction},
  author={Jureczko, Marian and Madeyski, Lech},
  booktitle={Proceedings of the 6th international conference on predictive models in software engineering},
  pages={1--10},
  year={2010}
}

@article{menzies2007data,
  title={Data mining static code attributes to learn defect predictors},
  author={Menzies, Tim and Greenwald, Jeremy and Frank, Art},
  journal={IEEE transactions on software engineering},
  volume={33},
  number={1},
  pages={2--13},
  year={2007},
  publisher={IEEE}
}

@article{lessmann2008benchmarking,
  title={Benchmarking classification models for software defect prediction: A proposed framework and novel findings},
  author={Lessmann, Stefan and Baesens, Bart and Mues, Christophe and Pietsch, Swantje},
  journal={IEEE transactions on software engineering},
  volume={34},
  number={4},
  pages={485--496},
  year={2008},
  publisher={IEEE}
}

@inproceedings{dal2015calibrating,
  title={Calibrating Probability with Undersampling for Unbalanced Classification.},
  author={Dal Pozzolo, Andrea and Caelen, Olivier and Johnson, Reid A and Bontempi, Gianluca},
  booktitle={SSCI},
  pages={159--166},
  year={2015}
}

@article{gong2022comprehensive,
  title={A comprehensive investigation of the impact of class overlap on software defect prediction},
  author={Gong, Lina and Zhang, Haoxiang and Zhang, Jingxuan and Wei, Mingqiang and Huang, Zhiqiu},
  journal={IEEE transactions on software engineering},
  volume={49},
  number={4},
  pages={2440--2458},
  year={2022},
  publisher={IEEE}
}

@article{smith2014instance,
  title={An instance level analysis of data complexity},
  author={Smith, Michael R and Martinez, Tony and Giraud-Carrier, Christophe},
  journal={Machine learning},
  volume={95},
  number={2},
  pages={225--256},
  year={2014},
  publisher={Springer}
}

@article{lorena2019complex,
  title={How complex is your classification problem? a survey on measuring classification complexity},
  author={Lorena, Ana C and Garcia, Lu{\'\i}s PF and Lehmann, Jens and Souto, Marcilio CP and Ho, Tin Kam},
  journal={ACM Computing Surveys (CSUR)},
  volume={52},
  number={5},
  pages={1--34},
  year={2019},
  publisher={ACM New York, NY, USA}
}

@article{hullermeier2021aleatoric,
  title={Aleatoric and epistemic uncertainty in machine learning: An introduction to concepts and methods},
  author={H{\"u}llermeier, Eyke and Waegeman, Willem},
  journal={Machine learning},
  volume={110},
  number={3},
  pages={457--506},
  year={2021},
  publisher={Springer}
}

@article{kuncheva2003measures,
  title={Measures of diversity in classifier ensembles and their relationship with the ensemble accuracy},
  author={Kuncheva, Ludmila I and Whitaker, Christopher J},
  journal={Machine learning},
  volume={51},
  number={2},
  pages={181--207},
  year={2003},
  publisher={Springer}
}

@article{bowes2018software,
  title={Software defect prediction: do different classifiers find the same defects?},
  author={Bowes, David and Hall, Tracy and Petri{\'c}, Jean},
  journal={Software Quality Journal},
  volume={26},
  number={2},
  pages={525--552},
  year={2018},
  publisher={Springer}
}

@inproceedings{moser2008comparative,
  title={A comparative analysis of the efficiency of change metrics and static code attributes for defect prediction},
  author={Moser, Raimund and Pedrycz, Witold and Succi, Giancarlo},
  booktitle={Proceedings of the 30th international conference on Software engineering},
  pages={181--190},
  year={2008}
}

@inproceedings{nagappan2005use,
  title={Use of relative code churn measures to predict system defect density},
  author={Nagappan, Nachiappan and Ball, Thomas},
  booktitle={Proceedings of the 27th international conference on Software engineering},
  pages={284--292},
  year={2005}
}

@inproceedings{zimmermann2009cross,
  title={Cross-project defect prediction: a large scale experiment on data vs. domain vs. process},
  author={Zimmermann, Thomas and Nagappan, Nachiappan and Gall, Harald and Giger, Emanuel and Murphy, Brendan},
  booktitle={Proceedings of the 7th joint meeting of the European software engineering conference and the ACM SIGSOFT symposium on The foundations of software engineering},
  pages={91--100},
  year={2009}
}

@inproceedings{nam2013transfer,
  title={Transfer defect learning},
  author={Nam, Jaechang and Pan, Sinno Jialin and Kim, Sunghun},
  booktitle={2013 35th international conference on software engineering (ICSE)},
  pages={382--391},
  year={2013},
  organization={IEEE}
}

@article{ma2021test,
  title={Test selection for deep learning systems},
  author={Ma, Wei and Papadakis, Mike and Tsakmalis, Anestis and Cordy, Maxime and Traon, Yves Le},
  journal={ACM Transactions on Software Engineering and Methodology (TOSEM)},
  volume={30},
  number={2},
  pages={1--22},
  year={2021},
  publisher={ACM New York, NY, USA}
}

@inproceedings{yang2025revisiting,
  title={Revisiting unnaturalness for automated program repair in the era of large language models},
  author={Yang, Aidan ZH and Kolak, Sophia and Hellendoorn, Vincent and Martins, Ruben and Le Goues, Claire},
  booktitle={2025 IEEE/ACM 47th International Conference on Software Engineering (ICSE)},
  pages={2561--2573},
  year={2025},
  organization={IEEE}
}

@article{wang2013using,
  title={Using class imbalance learning for software defect prediction},
  author={Wang, Shuo and Yao, Xin},
  journal={IEEE Transactions on Reliability},
  volume={62},
  number={2},
  pages={434--443},
  year={2013},
  publisher={IEEE}
}

@article{laradji2015software,
  title={Software defect prediction using ensemble learning on selected features},
  author={Laradji, Issam H and Alshayeb, Mohammad and Ghouti, Lahouari},
  journal={Information and Software Technology},
  volume={58},
  pages={388--402},
  year={2015},
  publisher={Elsevier}
}

@inproceedings{gal2016dropout,
  title={Dropout as a bayesian approximation: Representing model uncertainty in deep learning},
  author={Gal, Yarin and Ghahramani, Zoubin},
  booktitle={international conference on machine learning},
  pages={1050--1059},
  year={2016},
  organization={PMLR}
}

@article{ozturk2017type,
  title={Which type of metrics are useful to deal with class imbalance in software defect prediction?},
  author={{\"O}zt{\"u}rk, Muhammed Maruf},
  journal={Information and Software Technology},
  volume={92},
  pages={17--29},
  year={2017},
  publisher={Elsevier}
}

@article{van2022harm,
  title={The harm of class imbalance corrections for risk prediction models: illustration and simulation using logistic regression},
  author={Van den Goorbergh, Ruben and Van Smeden, Maarten and Timmerman, Dirk and Van Calster, Ben},
  journal={Journal of the American Medical Informatics Association},
  volume={29},
  number={9},
  pages={1525--1534},
  year={2022},
  publisher={Oxford University Press}
}

@article{ovadia2019can,
  title={Can you trust your model's uncertainty? evaluating predictive uncertainty under dataset shift},
  author={Ovadia, Yaniv and Fertig, Emily and Ren, Jie and Nado, Zachary and Sculley, David and Nowozin, Sebastian and Dillon, Joshua and Lakshminarayanan, Balaji and Snoek, Jasper},
  journal={Advances in neural information processing systems},
  volume={32},
  year={2019}
}

@article{spearman1961proof,
  title={The proof and measurement of association between two things.},
  author={Spearman, Charles},
  year={1961},
  publisher={Appleton-Century-Crofts}
}

@article{kendall1938new,
  title={A new measure of rank correlation},
  author={Kendall, Maurice G},
  journal={Biometrika},
  volume={30},
  number={1-2},
  pages={81--93},
  year={1938},
  publisher={Oxford University Press}
}

@article{el2010foundations,
  title={On the Foundations of Noise-free Selective Classification.},
  author={El-Yaniv, Ran and others},
  journal={Journal of Machine Learning Research},
  volume={11},
  number={5},
  year={2010}
}

@inproceedings{shahini2024empirical,
  title={An Empirical Study on Just-in-time Conformal Defect Prediction},
  author={Shahini, Xhulja and Metzger, Andreas and Pohl, Klaus},
  booktitle={Proceedings of the 21st International Conference on Mining Software Repositories},
  pages={88--99},
  year={2024}
}

@inproceedings{shahini2025calibration,
  title={On the calibration of Just-in-time Defect Prediction},
  author={Shahini, Xhulja and Bartel, Jone and Pohl, Klaus},
  booktitle={2025 IEEE/ACM 22nd International Conference on Mining Software Repositories (MSR)},
  pages={14--26},
  year={2025},
  organization={IEEE}
}

@article{hendrycks2016baseline,
  title={A baseline for detecting misclassified and out-of-distribution examples in neural networks},
  author={Hendrycks, Dan and Gimpel, Kevin},
  journal={arXiv preprint arXiv:1610.02136},
  year={2016}
}

@inproceedings{lewis1995sequential,
  title={A sequential algorithm for training text classifiers: Corrigendum and additional data},
  author={Lewis, David D},
  booktitle={Acm sigir forum},
  volume={29},
  number={2},
  pages={13--19},
  year={1995},
  organization={ACM New York, NY, USA}
}

@inproceedings{scheffer2001active,
  title={Active hidden markov models for information extraction},
  author={Scheffer, Tobias and Decomain, Christian and Wrobel, Stefan},
  booktitle={International symposium on intelligent data analysis},
  pages={309--318},
  year={2001},
  organization={Springer}
}

@article{shannon1948mathematical,
  title={A mathematical theory of communication},
  author={Shannon, Claude Elwood},
  journal={The Bell system technical journal},
  volume={27},
  number={3},
  pages={379--423},
  year={1948},
  publisher={Nokia Bell Labs}
}

@article{ye2025software,
  title={Software Defect Prediction Model Based on AST and Deep Learning},
  author={Ye, Zezhi and Yu, Chenghai and Lu, Zhilong},
  journal={Scalable Computing: Practice and Experience},
  volume={26},
  number={5},
  pages={2183--2195},
  year={2025}
}

@article{qiu2025features,
  title={Features extraction and fusion by attention mechanism for software defect prediction},
  author={Qiu, Shaoming and E, Bicong and He, Jingjie},
  journal={PloS one},
  volume={20},
  number={4},
  pages={e0320808},
  year={2025},
  publisher={Public Library of Science San Francisco, CA USA}
}

@article{liu2024sedpgk,
  title={SeDPGK: Semi-supervised software defect prediction with graph representation learning and knowledge distillation},
  author={Liu, Wangshu and Yue, Ye and Chen, Xiang and Gu, Qing and Zhao, Pengzhan and Liu, Xuejun and Zhao, Jianjun},
  journal={Information and Software Technology},
  volume={174},
  pages={107510},
  year={2024},
  publisher={Elsevier}
}

@article{yu2024improving,
  title={Improving effort-aware defect prediction by directly learning to rank software modules},
  author={Yu, Xiao and Rao, Jiqing and Liu, Lei and Lin, Guancheng and Hu, Wenhua and Keung, Jacky Wai and Zhou, Junwei and Xiang, Jianwen},
  journal={Information and Software Technology},
  volume={165},
  pages={107250},
  year={2024},
  publisher={Elsevier}
}

@article{ju2025jit,
  title={JIT-CF: Integrating contrastive learning with feature fusion for enhanced just-in-time defect prediction},
  author={Ju, Xiaolin and Cao, Yi and Chen, Xiang and Gong, Lina and Chakma, Vaskar and Zhou, Xin},
  journal={Information and Software Technology},
  volume={182},
  pages={107706},
  year={2025},
  publisher={Elsevier}
}

@article{platt1999probabilistic,
  title={Probabilistic outputs for support vector machines and comparisons to regularized likelihood methods},
  author={Platt, John and others},
  journal={Advances in large margin classifiers},
  volume={10},
  number={3},
  pages={61--74},
  year={1999},
  publisher={Cambridge, MA}
}

@article{wu2004probability,
  title={Probability estimates for multi-class classification by pairwise coupling},
  author={Wu, Ting-Fan and Lin, Chih-Jen and Weng, Ruby C},
  journal={Journal of Machine Learning Research},
  volume={5},
  number={Aug},
  pages={975--1005},
  year={2004}
}

@article{gneiting2007strictly,
  title={Strictly proper scoring rules, prediction, and estimation},
  author={Gneiting, Tilmann and Raftery, Adrian E},
  journal={Journal of the American statistical Association},
  volume={102},
  number={477},
  pages={359--378},
  year={2007},
  publisher={Taylor \& Francis}
}

@inproceedings{nam2015heterogeneous,
  title={Heterogeneous defect prediction},
  author={Nam, Jaechang and Kim, Sunghun},
  booktitle={Proceedings of the 2015 10th joint meeting on foundations of software engineering},
  pages={508--519},
  year={2015}
}

@article{glenn1950verification,
  title={Verification of forecasts expressed in terms of probability},
  author={Glenn, W Brier and others},
  journal={Monthly weather review},
  volume={78},
  number={1},
  pages={1--3},
  year={1950},
  publisher={War Department, Office of the Chief Signal Officer}
}

@inproceedings{zadrozny2002transforming,
  title={Transforming classifier scores into accurate multiclass probability estimates},
  author={Zadrozny, Bianca and Elkan, Charles},
  booktitle={Proceedings of the eighth ACM SIGKDD international conference on Knowledge discovery and data mining},
  pages={694--699},
  year={2002}
}

@inproceedings{ghotra2015revisiting,
  title={Revisiting the impact of classification techniques on the performance of defect prediction models},
  author={Ghotra, Baljinder and McIntosh, Shane and Hassan, Ahmed E},
  booktitle={2015 IEEE/ACM 37th IEEE International Conference on Software Engineering},
  volume={1},
  pages={789--800},
  year={2015},
  organization={IEEE}
}

@inproceedings{kim2011dealing,
  title={Dealing with noise in defect prediction},
  author={Kim, Sunghun and Zhang, Hongyu and Wu, Rongxin and Gong, Liang},
  booktitle={Proceedings of the 33rd international conference on software engineering},
  pages={481--490},
  year={2011}
}

@inproceedings{weiss2022simple,
  title={Simple techniques work surprisingly well for neural network test prioritization and active learning (replicability study)},
  author={Weiss, Michael and Tonella, Paolo},
  booktitle={Proceedings of the 31st ACM SIGSOFT international symposium on software testing and analysis},
  pages={139--150},
  year={2022}
}

@inproceedings{xia2023automated,
  title={Automated program repair in the era of large pre-trained language models},
  author={Xia, Chunqiu Steven and Wei, Yuxiang and Zhang, Lingming},
  booktitle={2023 IEEE/ACM 45th International Conference on Software Engineering (ICSE)},
  pages={1482--1494},
  year={2023},
  organization={IEEE}
}

@article{schober2018correlation,
  title={Correlation coefficients: appropriate use and interpretation},
  author={Schober, Patrick and Boer, Christa and Schwarte, Lothar A},
  journal={Anesthesia \& analgesia},
  volume={126},
  number={5},
  pages={1763--1768},
  year={2018},
  publisher={LWW}
}
\end{document}